\documentclass[utf8]{aa}

\usepackage[varg]{txfonts}
\usepackage{graphicx} 
\usepackage{natbib}
\usepackage{amsmath, bm}  
\usepackage[colorlinks=true, citecolor=blue, linkcolor=blue]{hyperref}

\usepackage[switch]{lineno}

\begin{document}

   \title{Anisotropic diffusion of high-energy cosmic rays in magnetohydrodynamic turbulence}

   \author{Na-Na Gao
          \inst{1}
          \and
          Jian-Fu Zhang\inst{1}\fnmsep\inst{2}\fnmsep\inst{3}
          }

   \institute{Department of Physics, Xiangtan University, Xiangtan, Hunan 411105, China\\
   \email{jfzhang@xtu.edu.cn}
         \and
             Key Laboratory of Stars and Interstellar Medium, Xiangtan University, Xiangtan 411105, China
        \and
             Department of Astronomy and Space Science, Chungnam National University, Daejeon, Republic of Korea
             }

\abstract
{The origin of cosmic rays (CRs) and how they propagate remain unclear. Studying the propagation of CRs in magnetohydrodynamic (MHD) turbulence can help to comprehend many open issues related to CR origin and the role of turbulent magnetic fields.
}
{To comprehend the phenomenon of slow diffusion in the near-source region, we study the interactions of CRs with the ambient turbulent magnetic field to reveal their universal laws.
}
{We numerically study the interactions of CRs with the ambient turbulent magnetic field, considering pulsar wind nebula as a general research case. Taking the magnetization parameter and turbulence spectral index as free parameters, together with radiative losses, we perform three group simulations to analyze the CR spectral, spatial distributions, and possible CR diffusion types.
}
{Our studies demonstrate that (1) CR energy density decays with both its effective radius and kinetic energy in the form of power-law distributions; 
(2) the morphology of the CR spatial distribution strongly depends on the properties of magnetic turbulence and the viewing angle; (3) CRs suffer a slow diffusion near the source and a fast/normal diffusion away from the source; (4) the existence of a power-law relationship between the averaged CR energy density and the magnetization parameter is independent of both CR energy and radiative losses; (5) radiative losses can suppress CR anisotropic diffusion and soften the power-law distribution of CR energy density.
}
{The distribution law established between turbulent magnetic fields and CRs presents an intrinsic property, providing a convenient way to understand complex astrophysical processes related to turbulence cascades.
}


\maketitle
%

\section{Introduction}

Cosmic rays (CRs) are charged particles moving with relativistic speeds, which can be directly detected in and near the solar system and indirectly observed from the emissions they produce far from the solar system (such as in the Milky Way and another galactic disk; \citealt{Armillotta2022}). Exploring the propagation processes of the Galactic CRs, including advection, diffusion, scattering, acceleration/reacceleration, and energy loss processes, will help to understand their origins (\citealt{Castellina2005, Strong2007, Blasi2013, Grenier2015}). Since propagating CRs will inevitably interact with turbulent magnetic fields, it also provides a useful tool for probing the properties of the interstellar medium (ISM). 

Magnetohydrodynamic (MHD) turbulence is ubiquitous in astrophysical plasmas ranging from interplanetary space to interstellar and intergalactic media. It has been pointed out that MHD turbulence directly or indirectly affects the propagation of CRs. Studying the propagation of CRs in turbulent magnetic fields can help us to comprehend many important issues both in space and astrophysics, e.g., solar modulation of CRs (\citealt{Jokipii1969}), driving Galactic winds (\citealt{Wiener2017, Krumholz2020}), CR anisotropy (\citealt{Qiao2023, Li2024}), X-rays binaries (\citealt{Zhang2018}), diffuse $\gamma-$ray emission (\citealt{Yan2012}), as well as the confinement and reacceleration of CRs (\citealt{Chandran2000, Yan2002, Zhang2021, Gao2024}).

The High-Altitude Water Cherenkov Observatory (HAWC) collaboration first reported the TeV $\gamma-$ray halos around the pulsars Geminga and Monogem (PSR B0656+14), with a spatial extension of $\sim$ 10-50 pc (\citealt{Abeysekara2017}). They explained these observations by the inverse Compton (IC) scattering process, that is, the high-energy CR electrons and positrons are injected from the pulsar wind nebula (PWN) and diffuse into the ISM to scatter off cosmic microwave background photons. \cite{Abeysekara2017} modeled the surface brightness profile of the $\gamma-$ray emission of these two TeV halos to constrain a low diffusion coefficient of $D=(4.5 \pm 1.2) \times 10^{27}\ {\rm cm}^2\ \rm s^{-1}$ at 100 TeV. This diffusion coefficient is hundreds of times lower than the average value in the Milky Way as inferred from the secondary-to-primary ratio measurements in the CR spectrum (i.e., the boron-to-carbon ratio (B/C); \citealt{Aguilar2016}). This suggests that the diffusion coefficient may be highly inhomogeneous on a small scale. It could be meaningful to investigate the origin of this slow-diffusion region around the pulsars, which can help us to understand many important issues, such as the CR position excess (\citealt{Hooper2017, Fang2018, Fang2019b, Tang2019, Xi2019, Manconi2020}), diffusion TeV $\gamma-$ray excess (\citealt{Linden2018, Yan2024}), and particle propagation near CR source (\citealt{Recchia2021, Lopez-Coto2022}). 

At present, there are mainly three plausible explanations for this slow diffusion as follows. Firstly, this slow diffusion is explained by the projection effects under the framework of anisotropic diffusion, which requires the nearby turbulent environment to be sub-Alfv\'enic and a small viewing angle between the local mean magnetic field and the line of sight (LOS) (\citealt{Liu2019, De2022, Fang2023}). Secondly, it may be caused by the turbulent environment generated by the parent supernova remnants of the pulsars (\citealt{Fang2019a}) or the escaped electrons themselves (\citealt{Evoli2018, Mukhopadhyay2022}). Alternatively, the mirror diffusion\footnote{\cite{Lazarian2021} found that in MHD turbulence CRs would not be trapped, due to the perpendicular super-diffusion of turbulent magnetic fields (\citealt{Xu2013, Lazarian2014, Hu2022}). On the contrary, they bounce between different magnetic mirrors and move along the local magnetic field, which leads to a new diffusion mechanism called mirror diffusion. With the stochastic change of pitch angles due to gyroresonant scattering, CRs undergo slow mirror diffusion at relatively large pitch angles and fast scattering diffusion at smaller pitch angles, resulting in a Levy-flight-like propagation that is slower than that induced by scattering alone (\citealt{Xu2021, Zhangchao2023, Barreto-Mota2024, Xiao2024}). As estimated in \cite{Barreto-Mota2024}, the physical value of the diffusion coefficient is $D_{\parallel} \leq 10^{27}\ {\rm cm^2\ s^{-1}}$ for 100 TeV CRs within an extended zone with tens of pc, which is consistent with the expected value to explain the observations of extended sources such as Geminga and Monogem (\citealt{Abeysekara2017}).} may be a channel for understanding this slow diffusion process because the combination of resonance scattering with mirror diffusion can cause a slower diffusion than that induced by scattering alone (\citealt{Lazarian2021, Xu2021, Zhangchao2023, Barreto-Mota2024}). 

Due to the presence of ubiquitous magnetic turbulence, the resulting CR anisotropic diffusion may be reasonable in the key astrophysical environments (\citealt{Holod2005, Parrish2005, Parrish2008}) and has obtained theoretical support (\citealt{Evoli2012, Cerri2021}). Recently, to explain the slow diffusion coefficient, the size, and morphology of observed TeV halos, \cite{Liu2019} and \cite{De2022} limited the Alfv\'en Mach number to a small value of $M_{\rm A} \lesssim 0.3$, and the viewing angle to $\psi \lesssim 5^{\circ}$. To some extent, these two small phase spaces may impede the application of the anisotropic diffusion model.     

The main motivations for the current work are the need for slow diffusion to explain the observations mentioned above and for the narrow parameter space constraints in the anisotropic diffusion model. Therefore, in the framework of CR anisotropic diffusion, we explore the interactions of high-energy electrons with the ambient turbulent magnetic fields, considering PWN, i.e., the Crab-like nebula, as a general research case. We want to know whether the slow-diffusion phenomenon could happen in a reasonable parameter space. The relevant CR diffusion law related to MHD turbulence theory to what extent is applicable for understanding the transport of CR particles.   

This paper is organized as follows. In Sect. \ref{sec:theor}, we briefly introduce the theoretical description of diffusion properties of CRs in MHD turbulence, including the energy-dependent parallel diffusion and anisotropic diffusion model. We perform a parameterized simulation and describe our initial simulation setup in Sect. \ref{sec:numre.Model}. We present the numerical results in Sect. \ref{sec:numre.Result}. Discussion and summary are provided in Sects. \ref{sec:discu} and \ref{sec:summary}, respectively.

\section{Theoretical basis}\label{sec:theor}
The propagation process of CRs can be described with the Fokker-Planck equation (\citealt{Gaisser1990, Berezinskii1990, Schlickeiser2002}), which is defined by
\begin{equation}
\begin{aligned}
  \frac{\partial f}{\partial t} = &\frac{1}{r_{\perp}}\frac{\partial}{\partial r_{\perp}}(r_{\perp} D_{\perp}  \frac{\partial f}{\partial r_{\perp}}) + D_{\parallel} (\frac{\partial^2 f}{\partial r_{\parallel}^2}) \\
  &+ \frac{1}{E^2} \frac{\partial}{\partial E}(E^2 D_{\rm EE} \frac{\partial f}{\partial E}) + \frac{\partial}{\partial E}(\dot{E} f) + S \,, \label{eq:FPeq}
\end{aligned}
\end{equation}
where $f(\bm{r},E,t)$ is the CR distribution function in the position space $\bm{r} = (x,\ y,\ z)$ and the kinetic energy space $E$. $r_{\parallel} = x$ is the coordinate along the mean magnetic field, and $r_{\perp} = \sqrt{y^2 + z^2}$ is the coordinate perpendicular to the mean magnetic field. $D_{\parallel}$ and $D_{\perp}$ are the spatial diffusion coefficients parallel to the mean magnetic field and perpendicular to it, respectively. $D_{\rm EE} = E^{2}v_{\rm A}^{2}/9D_{\parallel}$ is the energy diffusion coefficient (\citealt{Michalek1996}), and $\dot{E}$ is the rate of energy loss. In the case of power-law injection, the source function is defined by $S = \frac{d\dot{n}}{dE}\delta(\bm{r})$, where $d\dot{n}/dE \propto E^{-q}$ with a power-law index $q$ over the range from $E_0$ to $E_1$. While for mono-energy injection, the source function is defined by $S = \frac{\mathcal{L}}{E}\delta(\bm{r})\delta(E-E_{\rm in})$, where $\mathcal{L} = 10^{38}\ {\rm erg\ s^{-1}}$, $E_{\rm in}$, and $m_{\rm e}$ are the source luminosity, initial energy, and mass of the injected electrons, respectively.

\subsection{The effect of the scaling properties of MHD turbulence on the spatial diffusion}\label{sec:theor.diffturbulence}
It has been claimed that the propagation of CRs in the parallel direction with respect to the mean magnetic field is associated with the spectral properties of their ambient turbulent magnetic fields. In general, the parallel diffusion coefficient characterizing CR propagation can be described by the following empirical relation associated with the CR kinetic energy (\citealt{Seo1994, Trotta2011, Dempers2020}) 
\begin{equation}
  D_{\parallel} = D_0(E/m_{\rm e}c^2)^{\delta} \,, \label{eq:Dpar}
\end{equation}
where $c$ and $\delta$ are the light speed and diffusion index, respectively. The normalized spatial diffusion coefficient is set to be the typical ISM diffusion value, $D_0 = 1.0 \times 10^{28}\ {\rm cm^2\ s^{-1}}$, throughout the work (\citealt{Heesen2019}). It was claimed that the diffusion index of $\delta \in $ [0.3, 0.6] could match CR observations in the Milky Way remarkably well (\citealt{Strong2007, Trotta2011, Gaggero2014, Hopkins2022}).

The spectral energy density of interstellar turbulence presents a power-law relationship of $w(k)dk \sim k^{-\gamma}dk$, where $k$ and $\gamma = 2-\delta$ are the wave number and turbulence spectral index, respectively. The power-law index $\gamma = 5/3$ (known as a Kolmogorov spectrum \cite{Kolmogorov1941}; i.e., $\delta=$ 1/3) over a wide range of wave numbers $1/(10^{20}\ {\rm cm}) < k < 1/(10^{8}\ {\rm cm})$ (\citealt{Elmegreen2004}) determines the scaling law of the energy-dependent parallel diffusion coefficient $D_{\parallel} \sim E^{1/3}$. Theoretically, the Kolmogorov-type spectrum may refer only to a part of the MHD turbulence that includes the anisotropic (Alfv\'enic or slow modes) structures strongly elongated the mean magnetic field direction (\citealt{Goldreich1995}, hereafter GS95; \citealt{Cho2002}). On the other hand, there may exist another isotropic (fast mode) part of the turbulence in the ISM (\citealt{Cho2002}), with the exponent $\gamma = 3/2$ (i.e., $\delta$ = 1/2) typical for the Kraichnan-type turbulence spectrum (\citealt{Kraichnan1965}), in which the scaling of $D_{\parallel} \sim E^{1/2}$ is close to the high-energy asymptotic form of the diffusion coefficient obtained in the plain diffusion version of the empirical propagation model. In addition, in the case of Bohm-type diffusion, the diffusion index $\delta$ is typically assumed to be 1 (\citealt{Caballero-Lopez2004, Kempski2022, Lu2023}).

\subsection{The effect of the magnetization parameter on the spatial diffusion}\label{sec:theor.diffMA}
The propagation of CRs is subject to their interactions with turbulent magnetic fields. As expected, the ratio of the perpendicular diffusion coefficient to the parallel component should depend on the magnetization parameter, i.e., the Alfv\'en Mach number $M_{\rm A}$, with the power-law relationship of
\begin{equation}
  D_{\perp}/D_{\parallel} \approx M_{\rm A}^{\alpha} , \label{eq:DMa}
\end{equation}
which is the so-called anisotropic diffusion model.

In the case of the sub-Alfv\'enic turbulence, corresponding to a strong magnetic field, i.e., the field that cannot be easily bent at the turbulence injection scale $L_{\rm inj}$, individual magnetic field lines are aligned with the mean magnetic field. Note that \cite{Yan2008} stressed the importance of $M_{\rm A}^4$ dependence, in contrast to the $M_{\rm A}^2$ dependence in the classical studies (\citealt{Jokipii1969, Kota2000}). Later, analytical \citep{Lazarian2014} and numerical \citep{Xu2013,Maiti2022} results suggested a power-law index $\alpha \simeq$ 4 in both scales that are larger and less than $L_{\rm inj}$. 

For the super-Alfv\'enic turbulence, \cite{Lazarian2014} predicted a diffusion relation of the index $\alpha =$ 3 in the strong turbulence range of $[l_{\rm diss}, l_{\rm A}]$, where the $l_{\rm diss}$ and $l_{\rm A} = L_{\rm inj}/M_{\rm A}^3$ are the dissipation and transition scales, respectively. Namely, the dependence of $M_{\rm A}^3$ from CR diffusion reflects the scenario that the GS95 cascade starts at $l_{\rm A}$ rather than at the injection scale of turbulence $L_{\rm inj}$. This point has obtained support from simulations \citep{Maiti2022}. However, in the weak turbulence range from $l_{\rm A}$ to $L_{\rm inj}$, it has been predicted as the power-law index $\alpha =$ 0 \citep{Lazarian2014}, which means the diffusion is isotropic. In other words, the perpendicular diffusion coefficient is the same as the parallel one, i.e., $D_{\perp}\sim D_{\parallel}$. This point still needs a numerical test.

\section{Numerical setup}\label{sec:numre.Model}
To explore the interactions of high-energy particles with ambient magnetic turbulence, we numerically solve Eq. (\ref{eq:FPeq}) using the CR transport code CRIPTIC (\citealt{Krumholz2022}). With a PWN environment, we consider a uniform region containing gas with a density of $\rho \simeq 10^{-24}\ {\rm g\ cm^{-3}}$ and velocity of $v \simeq$ 300 $\rm km\ s^{-1}$, threaded by a uniform magnetic field $B_0$ in the $x$-axis direction. For the general environment of PWN, we have the typical temperature of thermal gas of $T\simeq 10^{5}$ K (\citealt{Zyuzin2021}) and the sonic speed of $c_{\rm s} \simeq$ 200 $\rm km\ s^{-1}$, resulting in a slightly supersonic Mach number of $M_{\rm s} = v/c_{\rm s} \simeq 1.5$. When involving radiative loss processes, we consider synchrotron radiative losses of relativistic electrons due to the presence of the ISM magnetic field and IC scattering losses due to both interstellar radiation field with a temperature of 20 K and cosmic microwave one in the Klein-Nishina regime.

With the purpose of exploring how the magnetization parameter $M_{\rm A}$ and turbulence spectral index $\gamma$ affect particles' diffusion behavior\footnote{In practice, we control the value of $M_{\rm A}$ by regulating the turbulent magnetic field strength $B_0$. For example, setting $B_0= 210\ \rm \mu G$, one has $M_{\rm A}=0.51$. }, together with various radiative processes, we divide our simulations into three groups:

\begin{itemize}
\item ${\rm Group}\ \mathcal{A}$: {\it the influence of radiative losses on CR diffusion}. In this case, with the fixed $\gamma=$ 5/3 and $M_{\rm A}=0.51$, we change the energy range and distribution of CR electrons. We consider two distribution types of CR electrons: a power-law energy distribution of $d\dot{n}/dE \propto E^{-q}$ with $q=$ 2.2 over the range from $E_0$ to $E_1$, and another mono-energy distribution.

\item ${\rm Group}\ \mathcal{B}$: {\it the influence of turbulence spectral index on CR diffusion}. With the fixed $M_{\rm A}=0.51$ and a series of mono-energy distributions, we change the spectral index of magnetic turbulence, corresponding to the different turbulence types. 

\item ${\rm Group}\ \mathcal{C}$: {\it the influence of the magnetization parameter on CR diffusion}. After fixing $\gamma=$ 5/3, we mainly consider the changes of the magnetization parameter $M_{\rm A} = v/v_{\rm A} = v/(B_0/\sqrt{4\pi \rho})$, which will match with different magnetic field strength. 
      
\end{itemize}

In the framework of the three groups described above, we run 90 sets of simulations up to $t=3\ {\rm kyr}$ with an output step of about 10 yr, considering the limited saving abilities of the data. Given a constant injection with the particles injection rate $\Gamma = 1.6\times 10^{2}\ \rm yr^{-1}$, the simulations include amounts of $4.8 \times 10^{5}$ particles. For more details, the interested reader can refer to Appendix A for numerical procedures.

\section{Numerical results}\label{sec:numre.Result}
\subsection{CR's spectral distribution} \label{sec:numre.CRspec}
As seen in Eq. (\ref{eq:FPeq}), the CR distribution function $f$ is a function of the position $\bm r$, the kinetic energy $E$, and the time $t$. Therefore, what we first explore is the behavior of the CR distributions associated with its energy density $U_{\rm CR} = E\ \int{f(r,E,t)\ dE}$ in units of $\rm GeV\ cm^{-3}$ (see \cite{Krumholz2022} for analytical solutions of $f$), where the effective radius of particles is defined as $r = \sqrt{(x/\sqrt{\chi})^2 + y^2 + z^2}$, with the factor $\chi = M_{\rm A}^{\alpha}$ corresponding to the anisotropy level between the perpendicular diffusion and the parallel one.{\footnote{The reason why we make the change of variables from $x$ to $x/\sqrt{\chi}$ is simply for the purposes of changing the form of the equation to one for which a standard Green’s function solution exists. Consequently, the energy density would be a function of radius and not related to angle. It should be noted that the diffusion radius is $R = \sqrt{x^2 + y^2 + z^2}$, which should be distinguished from the effective radius $r$.}}

\subsubsection{Effect of radiative losses on CR distributions} \label{sec:numre.CRspec.loss}

\begin{figure*}[t]
\centering
\includegraphics[width=0.48\textwidth,height=0.26\textheight]{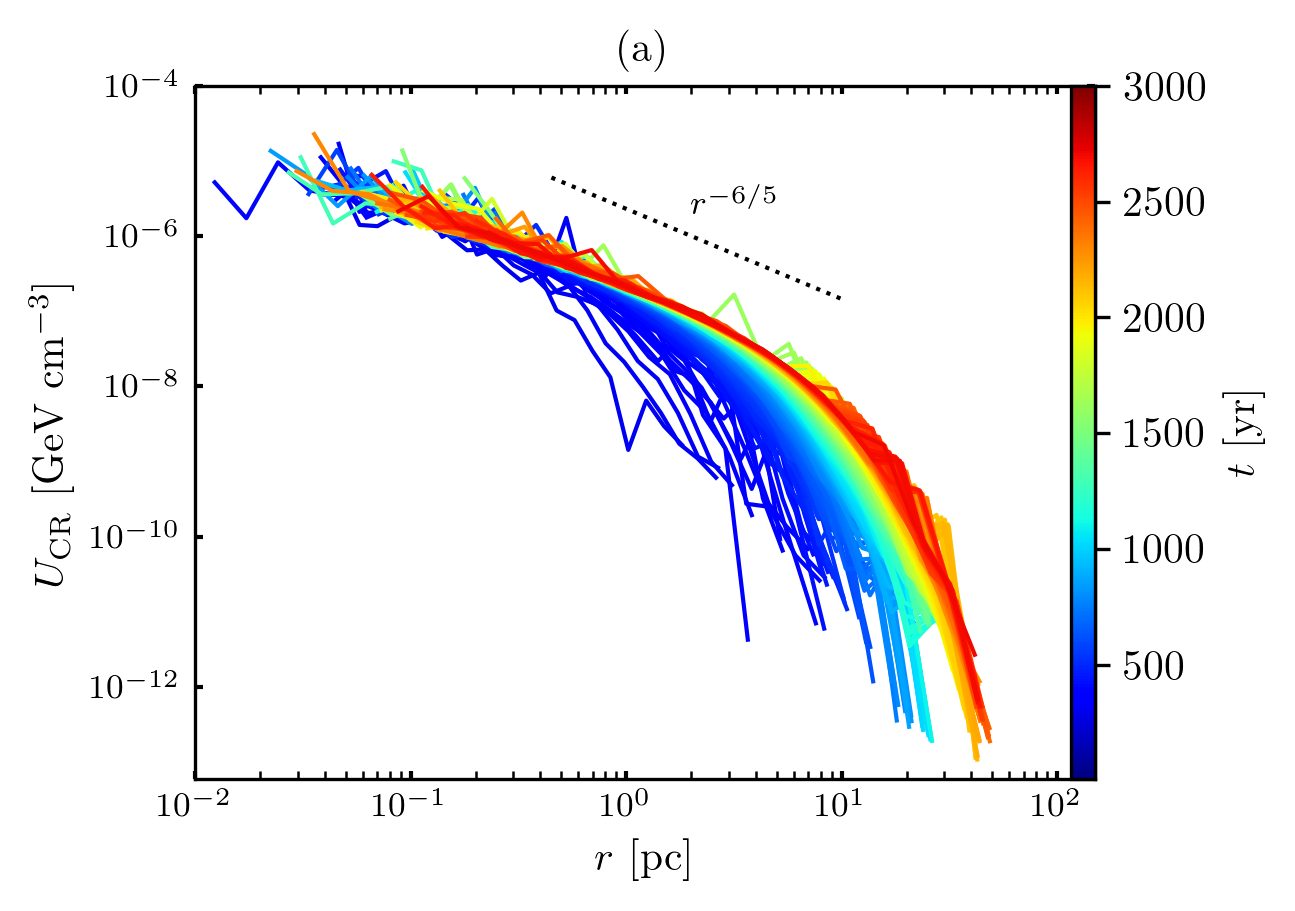}
\includegraphics[width=0.48\textwidth,height=0.26\textheight]{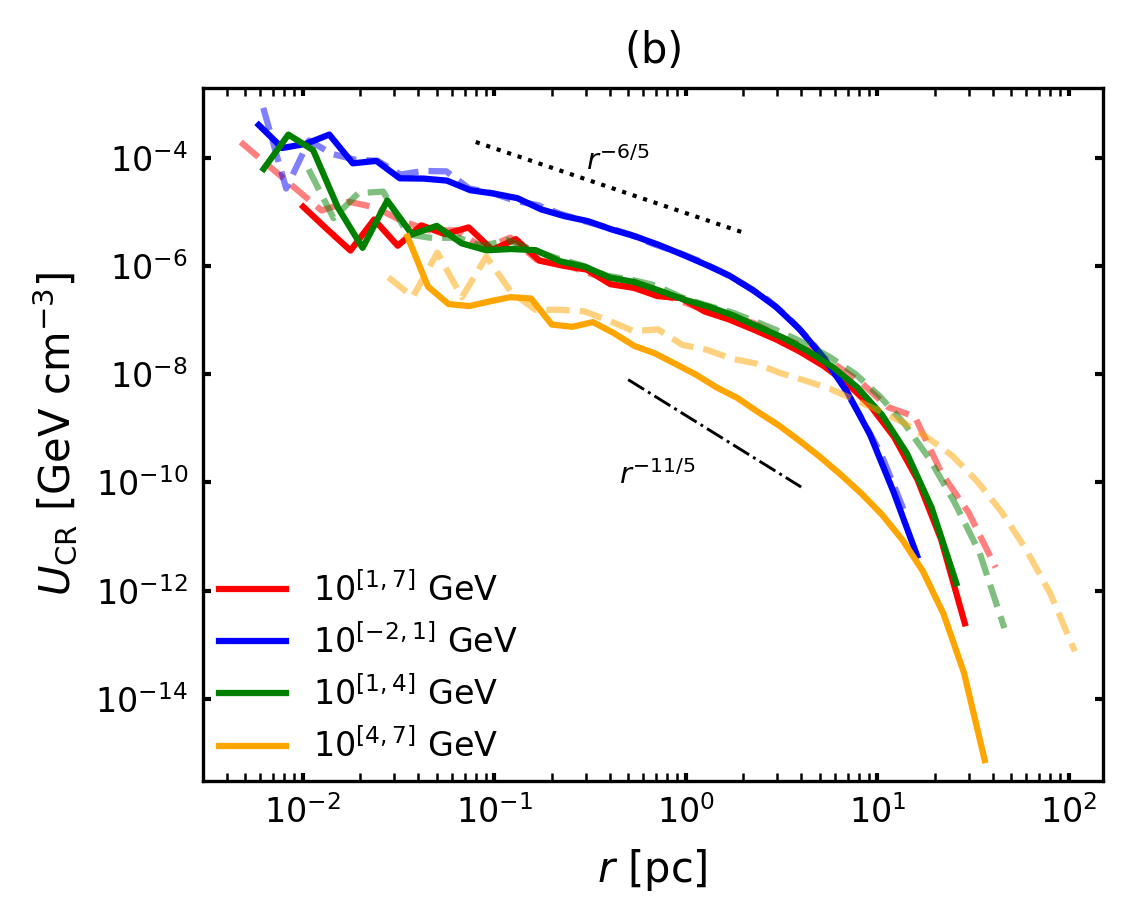}
\includegraphics[width=0.48\textwidth,height=0.26\textheight]{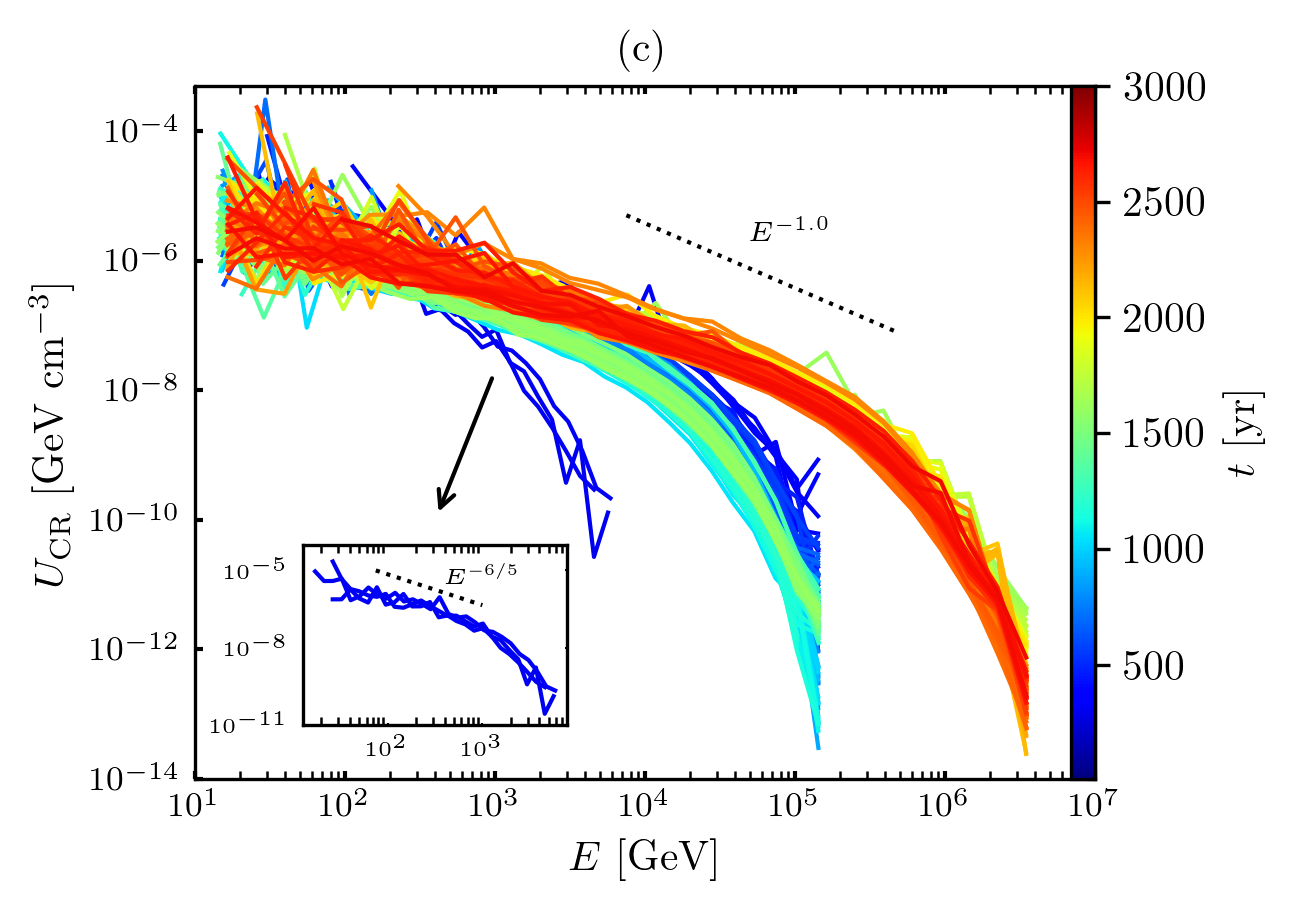}
\includegraphics[width=0.48\textwidth,height=0.26\textheight]{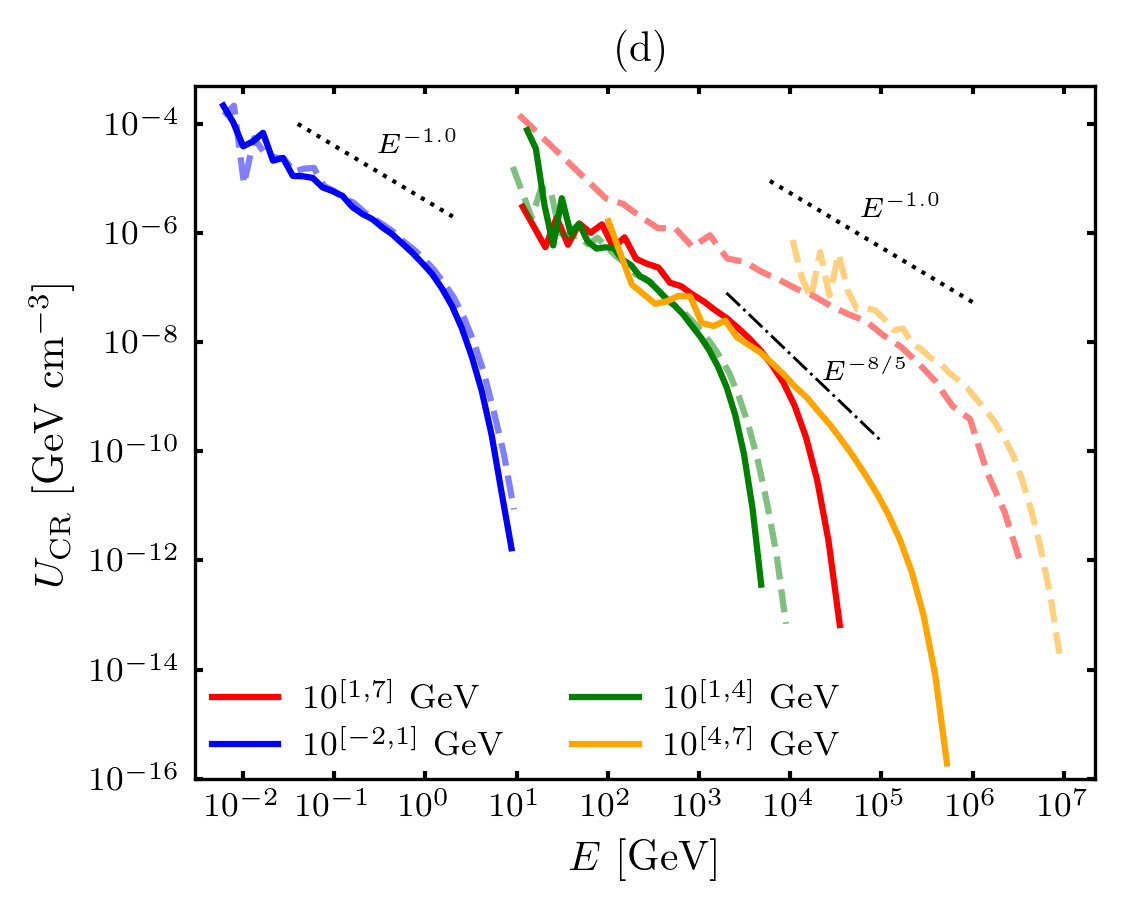}
\caption{The energy density $U_{\rm CR}$ as a function of the particles' effective radius $r$ (upper row) and kinetic energy $E$ (lower row). Left column: the time-dependent evolution of $U_{\rm CR}$ corresponds to the no loss case of $E\in 10^{[1, 7]}$ GeV, where the color bar shows the simulation time in units of yr. In panel (c), the $U_{\rm CR}$ distribution of the several initial snapshots is shown in the inset. Right column: $U_{\rm CR}$ distributions in four different energy regimes plotted at the final snapshot ($t=$ 3 kyr), where the dashed and solid lines show the results without and with loss processes, respectively. Simulations are based on the related parameters in Group $\mathcal{A}$.}
\label{Fig:01Ur_UE}
\end{figure*}

In the energy range from 1 GeV to 10 PeV, the energy density $U_{\rm CR}$ as a function of the effective radius and the evolution time is shown in Fig. \ref{Fig:01Ur_UE}(a) in the absence of radiative loss, from which can see that the time-dependent evolution of the energy density with the effective radius presents a power-law relationship of $U_{\rm CR} \propto r^{-6/5}$ near the source region and then cuts off exponentially. With time evolution, $U_{\rm CR}$ distributions almost follow the same power law with a gradually increasing range. After $t\sim 2.5$ kyr (orange-red curves), $U_{\rm CR}$ vs. $r$ reaches a statistical steady state. The fluctuations that appeared at different times are a result of the deviation from statistics. This result may imply that the CR electrons are suffering from slow diffusion near the source and fast/normal diffusion away from the source, which will help to understand observations of slow diffusion around the pulsar (\citealt{Abeysekara2017}).

To explore CR distributions from different energy particles, we divide the whole energy band into individual energy regimes: the GeV ($E\in 10^{[-2,\ 1]}$ GeV), TeV ($E\in 10^{[1,\ 4]}$ GeV), and PeV ($E\in 10^{[4,\ 7]}$ GeV) range, as well as a wide range from $10^1$ to $10^7$ GeV. Focusing on a comparison study with/without radiative loss processes, we provide the results in Fig. \ref{Fig:01Ur_UE}(b) obtained at the final snapshot. Without radiative losses, $U_{\rm CR}$ distributions at different energy ranges remain a relation of $U_{\rm CR} \propto r^{-6/5}$ around the source (about several or multi-tens pc) and then it enters the exponential decay stage. After including radiative losses, we see that with increasing the kinetic energy, the effect of loss processes gradually becomes stronger. In particular, the radiative losses modify the particle $U_{\rm CR}$ distribution in the PeV range, presenting a steeper power-law of $U_{\rm CR} \propto r^{-11/5}$. In this regard, the maximum effective radius the particles can reach is smaller, and the spatially extended power law range is narrower.
   
Based on the same simulation setup used in Fig. \ref{Fig:01Ur_UE}(a), we plot in Fig. \ref{Fig:01Ur_UE}(c) the evolution of energy density $U_{\rm CR}$ over the kinetic energy of CR electrons $E$ at the different evolution times. As is seen from this panel, $U_{\rm CR}$ presents a power-law relationship of $U_{\rm CR} \propto 1/E$ with an exponential high-energy cutoff. In addition, we also see a feature of the three discrete distributions of the energy density (rather than changing continuously): (1) the maximum energy of $E_{\rm max}\simeq 10^{3.7}$ GeV at about 10 yr; (2) $E_{\rm max}\simeq 10^{5.2}$ GeV from $\sim 40$ yr to $\sim 1.5\times 10^3$ yr, and (3) $E_{\rm max}\simeq 10^{6.5}$ GeV at multi-thousand yr. The power law relation of $U_{\rm CR} \propto 1/E$ we find is approximate to the initial injection form of $U_{\rm CR} \propto Ed\dot{n}/dE \propto E^{-6/5}$, i.e., a power-law energy injection of $d\dot{n}/dE \propto E^{-q}$ with $q=$ 2.2 in Group $\mathcal{A}$. Here, the slight hard power-law index feature should be a result of a weak stochastic acceleration process (associated with the third term of Eq. (\ref{eq:FPeq}) on the right-hand side). As a result, the evolution processes of CR electrons result in a modification of the power-law distribution, in particular the occurrence of a significant exponential cutoff.

Similar to Fig. \ref{Fig:01Ur_UE}(b), we show the distribution of $U_{\rm CR}$ vs. the energy $E$ in Fig. \ref{Fig:01Ur_UE}(d) at the final snapshot. In the absence of radiative losses, the CR electrons have a spectral energy distribution of $U_{\rm CR} \propto 1/E$, the slightly hardened spectral index of which is due to a weak acceleration process as explained in Fig. \ref{Fig:01Ur_UE}(c). When involving loss processes, the energy density presents a steeper relation of $U_{\rm CR} \propto E^{-8/5}$ with a significant exponential cutoff in the PeV range. This process gives rise to a change of $U_{\rm CR} \propto E^{-3/5}$ compared with the scenario without losses. However, when compared with the initial distribution of $U_{\rm CR} \propto E^{-6/5}$, there is a change of $U_{\rm CR} \propto E^{-2/5}$. This indicates that the diffusion processes are dominated by a strong radiative loss, resulting in the potential PeV observations (e.g., \citealt{Cao2021, Cao2024}). Furthermore, the particle distribution with a power law plus an exponential cutoff is suggested when understanding high-energy observations.

\subsubsection{Effect of MHD turbulence properties on CR distributions} \label{sec:numre.CRspec.turbulence}

\begin{figure*}[t]
\centering   \includegraphics[width=0.48\textwidth,height=0.24\textheight]{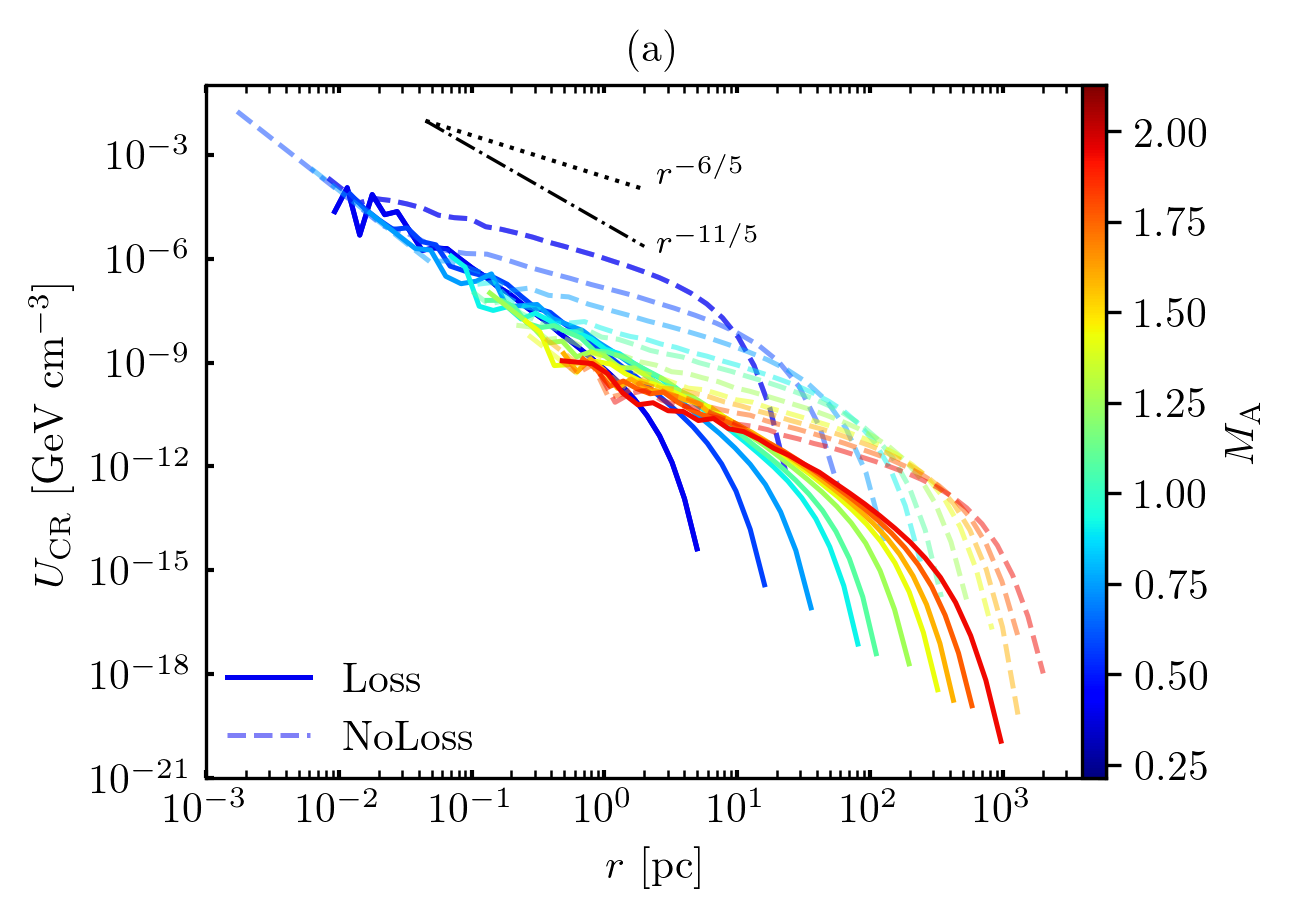}    \includegraphics[width=0.48\textwidth,height=0.24\textheight]{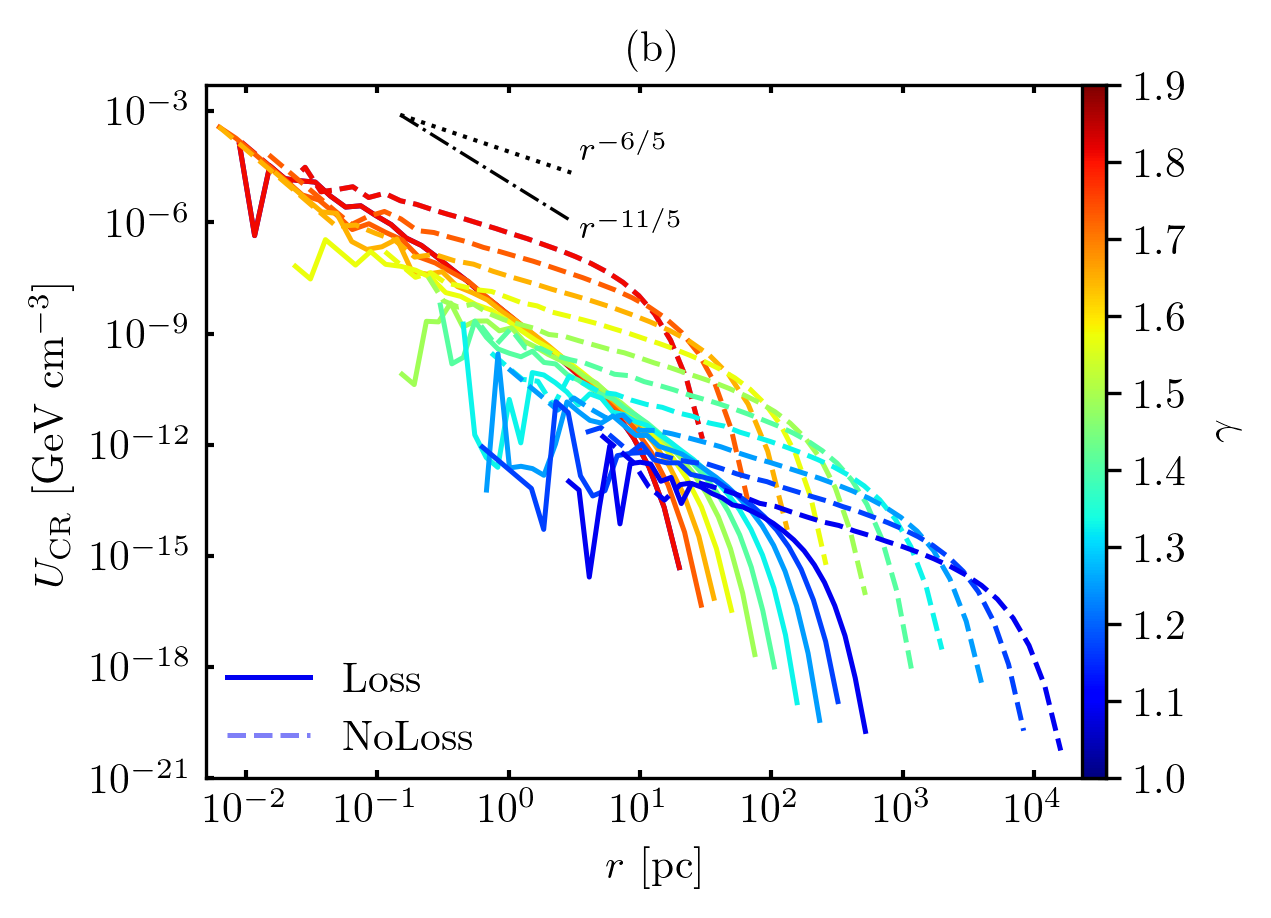}
\caption{Energy density $U_{\rm CR}$ of the mono-energy 1 PeV as a function of the particle's effective radius $r$. The results shown in panels (a) and (b) are based on Groups $\mathcal{C}$ and $\mathcal{B}$, respectively.}
    \label{Fig:02Ur_MA-Delta}%
\end{figure*}

Based on the final snapshot simulations from Group $\mathcal{C}$, we show the energy density $U_{\rm CR}$ as a function of the effective radius $r$ and the magnetization parameter $M_{\rm A}$ in Fig. \ref{Fig:02Ur_MA-Delta}(a). As studied in Sect. \ref{sec:numre.CRspec.loss}, given that the effect of the loss processes is significant only in the high-energy regime, at least PeV-order of magnitude, we here just show the results of mono-energy injection at $E = $ 1 PeV. The overall distribution of $U_{\rm CR}$ vs. $r$ in various $M_{\rm A}$ remains similar to that of the power law injection, i.e., presenting a power-law transition from $U_{\rm CR} \propto r^{-6/5}$ (without loss) to $U_{\rm CR} \propto r^{-11/5}$ (with loss). With increasing the magnetization parameter $M_{\rm A}$, the value of $U_{\rm CR}$ decreases, where CR electrons will diffuse away from the source and up to a larger spatial scale. This is because the larger $M_{\rm A}$ means a weaker magnetic field, i.e., weaker magnetic trapping effect, then the spatial extent caused by diffusion will be wider. Comparing the sub-Alfv\'enic ($M_{\rm A} < 1.0$) with super-Alfv\'enic turbulence regime ($M_{\rm A} > 1.0$), they are similar from the perspective of the overall power-law distribution. When focusing on the case of radiative losses (solid curves), we find that the increment of the maximum effective radius $\Delta r=r_{\rm M_{\rm A}=0.82}-r_{\rm M_{\rm A}=0.21} \approx 100$ pc for the sub-Alfv\'enic turbulence is smaller than that $\Delta r \approx 800$ pc for the super-Alfv\'enic turbulence. This may be due to the different dependence of the perpendicular diffusion coefficient on $M_{\rm A}$, i.e., $\alpha = 4.0$ for the sub-Alfv\'enic regime, while $\alpha = 3.0$ for the super-Alfv\'enic case (see Sect. \ref{sec:theor.diffMA}). This manifests that the magnetization parameter $M_{\rm A}$, is of great importance to the CR's spectral distribution. 

Fig. \ref{Fig:02Ur_MA-Delta}(b) shows the energy density $U_{\rm CR}$ as a function of the effective radius $r$ with various $\gamma$, based on the last simulation snapshot from Group $\mathcal{B}$. As shown, the power-law transition is similar to that of Fig. \ref{Fig:02Ur_MA-Delta}(a). With increasing the turbulence spectral index, the effective radius decreases while the energy density increases. We note that as $\gamma$ decreases, the fluctuation of the spectral distribution increases within several pc scales, which may be due to the effects of insufficient statistical samples and strong radiative losses. This demonstrates that the scaling of MHD turbulence is an important factor in the CR's spectral distribution.

\subsection{CR's anisotropy} \label{sec:numre.aniso}
\subsubsection{Effect of radiative losses on CR anisotropy} \label{sec:numre.aniso.loss}

\begin{figure*}[t]
\centering
\includegraphics[width=0.98\textwidth,height=0.24\textheight]{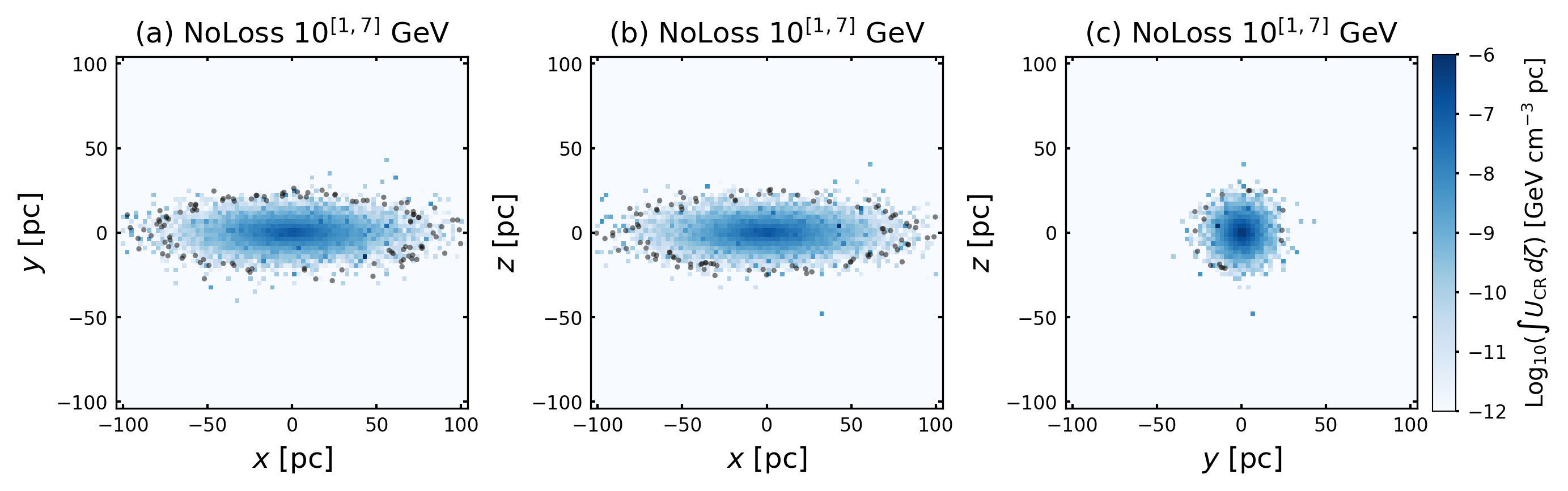}   
\includegraphics[width=0.98\textwidth,height=0.24\textheight]{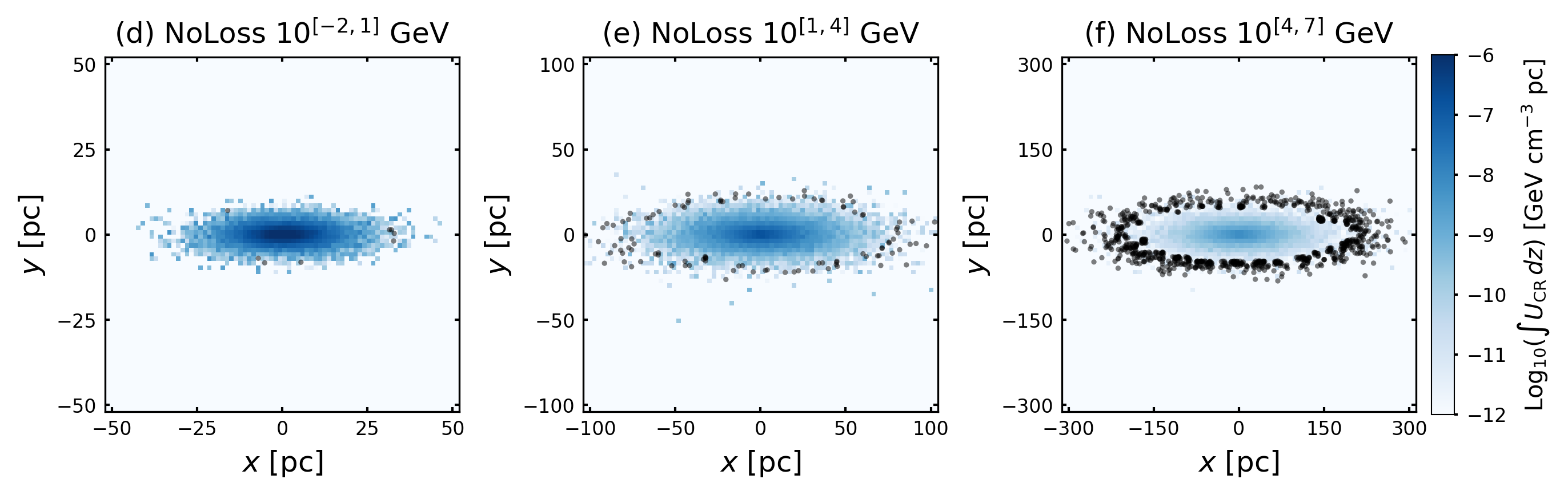}
\includegraphics[width=0.98\textwidth,height=0.24\textheight]{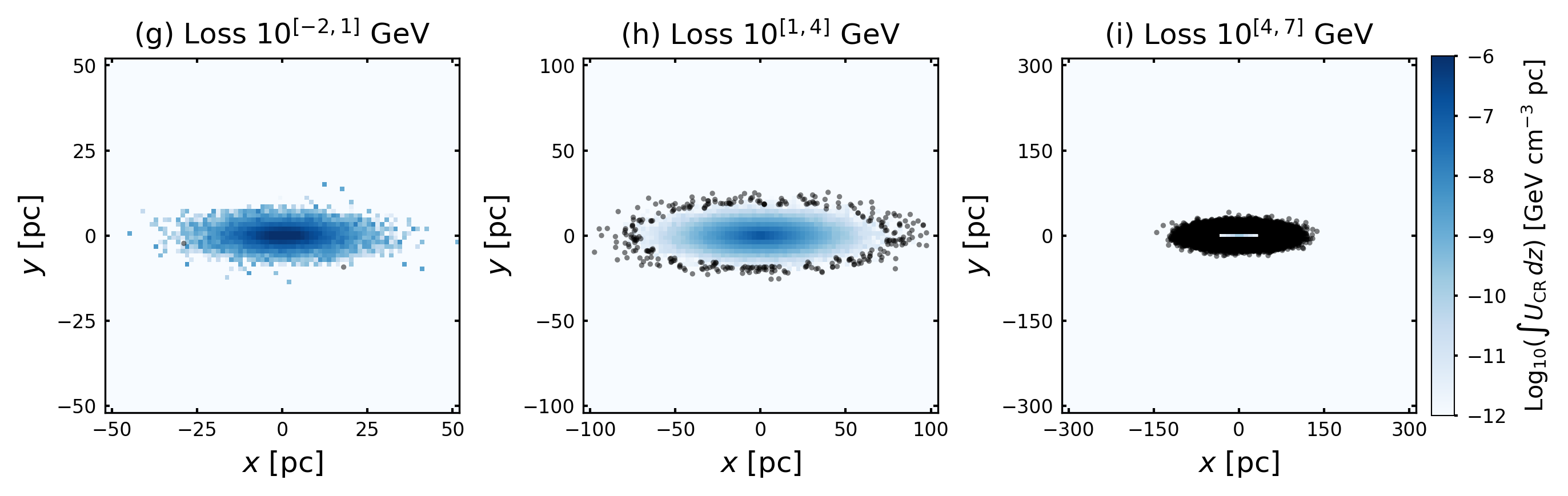}
\caption{Projected energy density $\int U_{\rm CR}\ d\zeta$ in $xy$, $xz$, and $yz$ planes. The energy density denoted by the color bars is limited in the range of $10^{-12} - 10^{-6}$ GeV cm$^{-3}$ pc. The black points show the positions of individual CR particles in the regions where the energy density falls below the range shown by the color bars.}
\label{Fig:03Anisotropy_E}%
\end{figure*}

We show the projected energy density $\int U_{\rm CR}\ d\zeta$ at the final snapshot for the power-law distribution of Group $\mathcal{A}$ in Fig. \ref{Fig:03Anisotropy_E}, including the non-loss case of $E\in 10^{[1,\ 7]}$ GeV projected in three different planes (upper row), the non-loss case of three narrow energy ranges projected in $xy$ plane (middle row), and the corresponding case with losses (lower row). As shown in the upper row of Fig. \ref{Fig:03Anisotropy_E}, the morphology is anisotropic and looks like ``ovals'', in $xy$ (panel a) and $xz$ (panel b) planes. While in the $yz$ plane (panel c), it is isotropic and spherically symmetrical. This indicates that the morphology is strongly affected by the viewing angle $\psi$. Given that the mean magnetic field is along with the $x$-axis direction, the morphology shows isotropic and spherically symmetrical at $\psi \sim 0^{\circ}$ (i.e., $yz$ plane), while it shows ``ovals'' at $\psi \sim 90^{\circ}$ (i.e., $xy$ and $xz$ planes). This is consistent with the previous results limiting $\psi \lesssim 5^{\circ}$ (\citealt{Liu2019,De2022}).

As shown in the middle and lower rows of Fig. \ref{Fig:03Anisotropy_E}, we can see that the morphology projected in $xy$ plane (i.e, $\psi \sim 90^{\circ}$) all show an anisotropy regardless of the case of high/low energy-band and loss/non-loss situation. Note that, for all cases considered here, the fixed Alfv\'en Mach number $M_{\rm A}=$ 0.51, corresponds to the sub-Alfv\'enic turbulence regime, with the initial ratio of $D_{\perp}/D_{\parallel} = 0.51^{4.0} \approx 0.07$ according to Eq. (\ref{eq:DMa}). As the energy increases, the spatial extent becomes wider, appearing a larger number of CR electrons with $\int U_{\rm CR}\ dz\lesssim 10^{-12}\ {\rm GeV\ cm^{-3}}$ pc (dark points). Comparing the non-loss with loss cases, we can see that the morphology is similar in both the GeV (left column) and TeV (middle column) bands. However, in the TeV band (middle column), the number of CR electrons with $\int U_{\rm CR}\ dz\lesssim 10^{-12}\ {\rm GeV\ cm^{-3}}$ pc for the loss case (panel h) is larger than that for the non-loss case (panel e). While for the PeV band (right column), the energy loss is so strong (see Fig. \ref{Fig:01Ur_UE}(d)) that the spatial extent decreases deeply and the number of low-energy particles increases (see the map filled with low-energy black points in Fig. \ref{Fig:03Anisotropy_E}(i)). As a result, the anisotropy of the particles' energy distribution originates from the anisotropy of MHD turbulence.

\subsubsection{Effect of MHD turbulence properties on CR anisotropy} \label{sec:numre.aniso.turbulence}

\begin{figure*}[t]
\centering
 \includegraphics[width=0.98\textwidth,height=0.24\textheight]{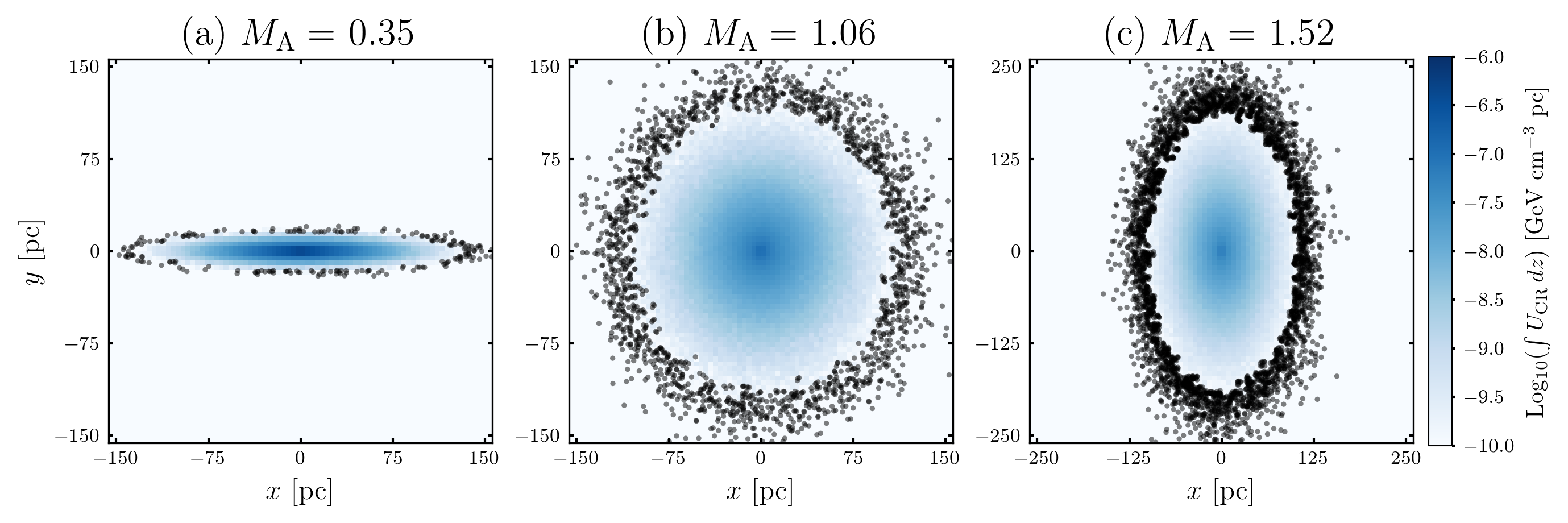}
 \includegraphics[width=0.98\textwidth,height=0.24\textheight]{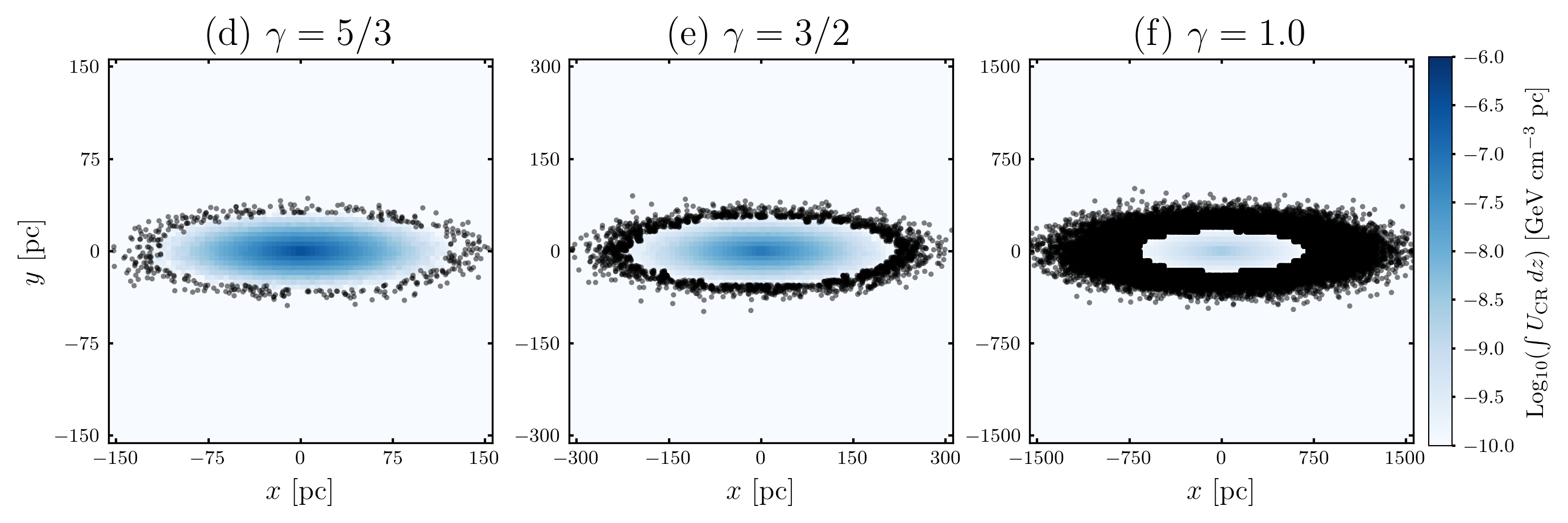}
\caption{Projected energy density $\int U_{\rm CR}\ dz$ in $xy$ plane for the case of $E= 1$ TeV including the various $M_{\rm A}$ (upper row, $\gamma=5/3$) and $\gamma$ (lower row, $M_{\rm A}=0.51$). The energy density denoted by the color bars is limited in the range of $10^{-10} - 10^{-6}$ GeV cm$^{-3}$ pc. The black points show the positions of individual CR particles in the regions where the energy density falls below the range shown by the color bars.}
\label{Fig:04Anisotropy_MA-Delta}%
\end{figure*}

Furthermore, we investigate the effect of $M_{\rm A}$ and $\gamma$, which are associated with the properties of MHD turbulence, on the anisotropic properties for the energy distribution of particles. As shown in upper row of Fig. \ref{Fig:04Anisotropy_MA-Delta}, in the case of $\gamma=$ 5/3, we plot the projected energy density $\int U_{\rm CR}\ dz$ in $xy$ plane including the magnetization parameters $M_{\rm A}=$ 0.35 (sub-Alfv\'enic turbulence; panel a), $M_{\rm A}=$ 1.06 (trans-Alfv\'enic; panel b), and $M_{\rm A}=$ 1.52 (super-Alfv\'enic; panel c). When $M_{\rm A}< 1$ (panel a), it can be seen that the spatial distribution of the projected energy density is highly anisotropic, and the spatial diffusion scale in the parallel direction ($x$ axis) is larger than that in the perpendicular direction ($y$ axis), which can be easily explained by Eq. (\ref{eq:DMa}) in the case of $\alpha=$ 4. In the framework of this anisotropic diffusion, the smaller the value of $M_{\rm A}$ is, the higher anisotropy. When $M_{\rm A}\simeq 1$ (panel b), the morphology becomes isotropic because of $D_{\perp}/D_{\parallel} \approx M_{A}^{\alpha} \approx 1$, meaning that the scale of the perpendicular diffusion approximates to the parallel diffusion one. When $M_{\rm A}>1$ (panel c), it is opposite to the case of the sub-Alfv\'enic regime, that is, the diffusion scale in the perpendicular direction ($y$ axis) being larger than that in the parallel direction ($x$ axis), and the larger value of $M_{\rm A}$ the higher anisotropy. Comparing these three maps, we see that with increasing the magnetization parameter, the spatial extent becomes larger in the $y$-axis direction, due to the increasing $D_{\perp}$. At the same time, we see that the number of black points is increasing, indicating that the number of CR particles in the regions where $U_{\rm CR}$ falls below the range shown in the color bars is increasing.

For the lower row of Fig. \ref{Fig:04Anisotropy_MA-Delta}, in the case of $M_{\rm A}=0.51$, we plot the projected energy density $\int U_{\rm CR}\ dz$ in $xy$ plane including the turbulence spectral indexes $\gamma =$ 5/3 (Kolmogorov turbulence; panel a), $\gamma =$ 1/2 (Kraichnan turbulence; panel b), and $\gamma =$ 1.0 (Bohm-type diffusion; panel c). In these cases, the anisotropy level of the spatial distribution of particles remains unchanged, based on the measurement of the aspect ratio in the $x$ and $y$ directions. We note that the spatial extent becomes larger with decreasing the turbulence spectral index. The reason why is that our simulation involves the energy-dependent diffusion model, as described by the parallel diffusion coefficient in Eq. (\ref{eq:Dpar}). At the same energy, the parallel diffusion coefficient $D_{\parallel}$ increases with decreasing $\gamma$, thus the scale of spatial diffusion becomes larger. With hardening the spectral index. i.e., decreasing $\gamma$, we see the phenomena of gradually increasing the number of CR particles, which should be related to the energy cascade of MHD turbulence. This indicates that the energy distributions of particles strongly depend on the spectral properties of the ambient turbulent magnetic fields.

\begin{figure}[t]
\centering
 \includegraphics[width=0.48\textwidth,height=0.32\textheight]{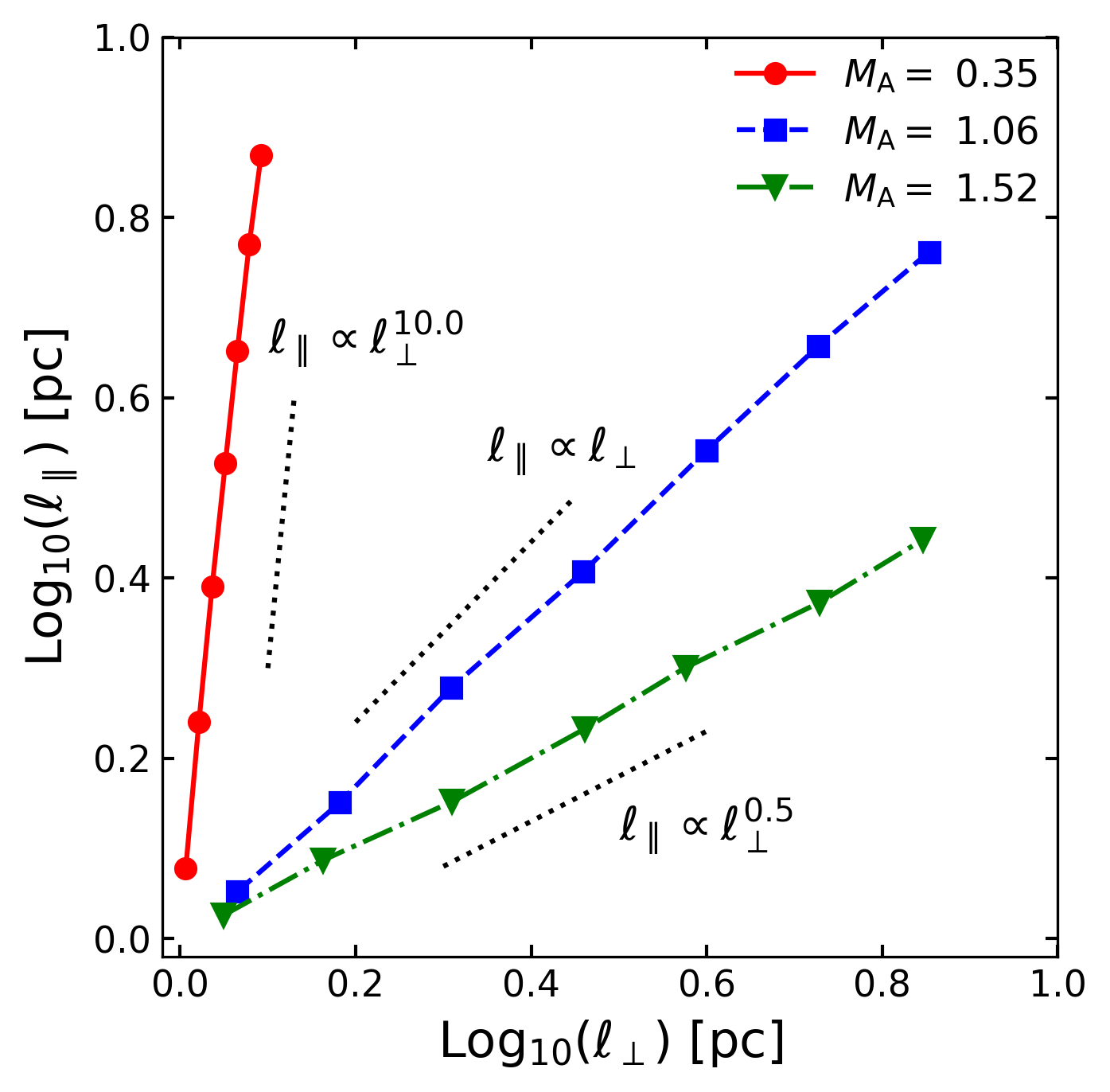}
\caption{Anisotropy scalings of the projected energy density $\int U_{\rm CR} dz$ in $xy$ plane, where the red-circle, blue-square, and green-triangle symbols are obtained by panels (a), (b), and (c) of Fig. \ref{Fig:04Anisotropy_MA-Delta}, respectively. Here, $l_{\parallel}$ and $l_{\perp}$ represent the parallel ($x$ axis) and perpendicular ($y$ axis) scales of the morphology, respectively.}
\label{Fig:08counter_anis}%
\end{figure}

To quantify the anisotropy of the spatial distribution of the projected energy density, we show its anisotropy scaling in Fig. \ref{Fig:08counter_anis}, corresponding to the three cases of the upper row from Fig. \ref{Fig:04Anisotropy_MA-Delta}. The results show that the anisotropy scaling satisfies $l_{\parallel}\propto l_{\perp}^{1/\eta}$, where the index $\eta=$ 0.1 (for $M_{\rm A}=0.35$), 1.0 ($M_{\rm A}=1.06$), and 2.0 ($M_{\rm A}=1.52$). Note that the anisotropy of the particle's spatial distribution deviates from the anisotropy relation of $l_{\parallel}\propto l_{\perp}^{2/3}$ to MHD turbulence. This is because the power-law index strongly depends on the magnetization parameter via a relation of $\eta=M_{\rm A}^{\alpha/2}$ derived by Eq. (\ref{eq:DMa}) together with the diffusion scale $l_{\parallel, \perp} \approx \sqrt{D_{\parallel, \perp}t}$ (\citealt{Liu2020}). This reflects that the particle's anisotropic distributions are mainly affected by the magnetic field strength, except for the effect of the viewing angle discussed above.

\subsection{CR's diffusion types} \label{sec:numre.aniso.diff}

\begin{figure}[t]
\centering   \includegraphics[width=0.48\textwidth,height=0.22\textheight]{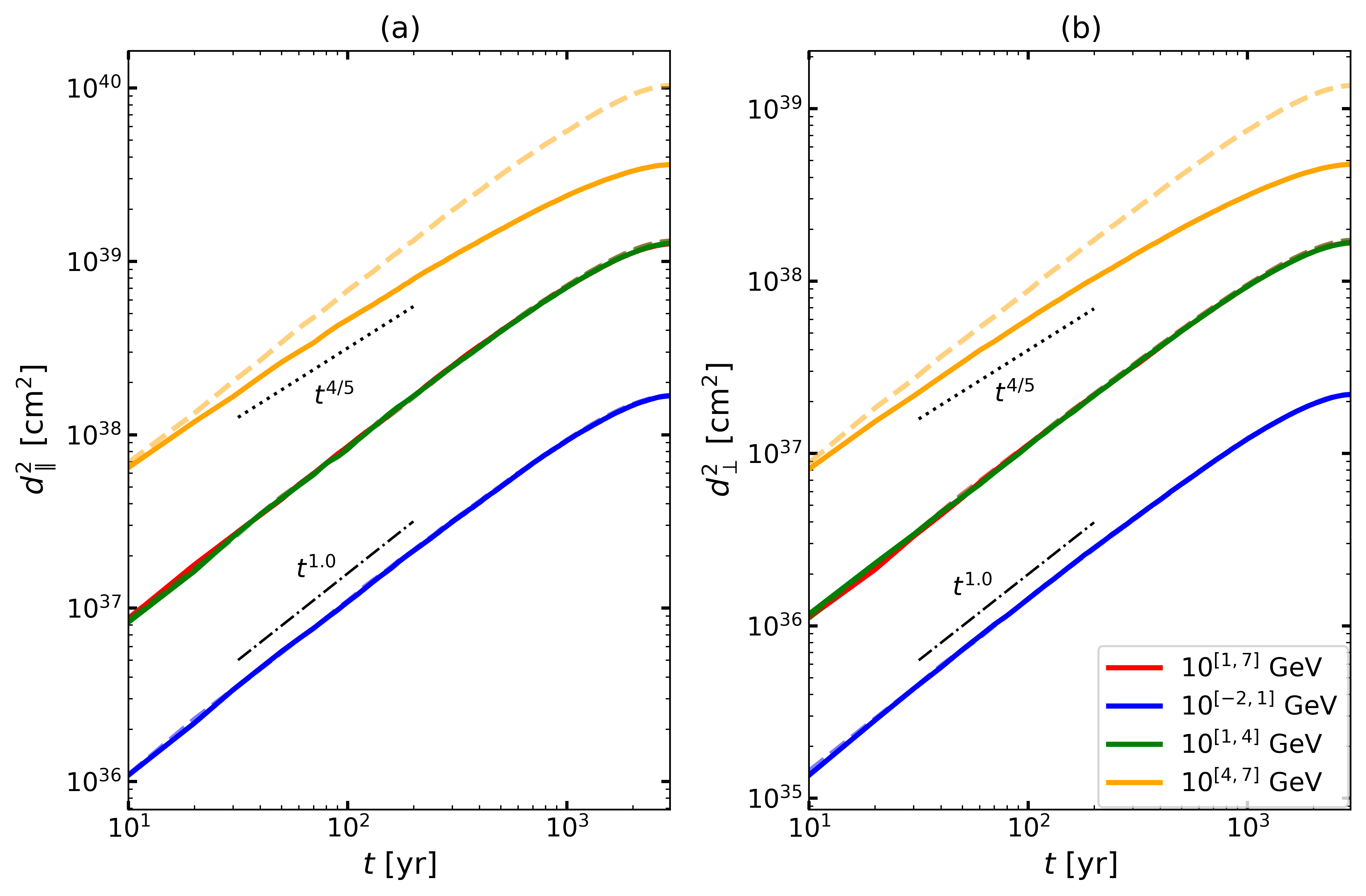}   
\caption{The parallel ensemble-averaged squared deviations of particles ($d_{\parallel}^2$; panel a) and the perpendicular one ($d_{\perp}^2$; panel b) as a function of the simulation time with (solid lines) and without (dashed lines) radiative loss processes. $d_{\parallel}^2$ (and $d_{\perp}^2$) of $E=10^{[1,\ 7]}$ GeV (red lines) and $E=10^{[1,\ 4]}$ GeV (green lines) overlap together.}
\label{Fig:05Dt_E}
\end{figure}

To understand the spatial diffusive behavior of particles, we show the evolution of the ensemble-averaged squared deviations{\footnote{Given $d_{\parallel,\perp}^2 \propto t^{\nu}$, when $\nu<1.0$, it is called sub-diffusion; when $\nu=1.0$, it is called normal-diffusion; when $\nu>1.0$, it is called super-diffusion (\citealt{Ostrowski1997, Sioulas2020}).}} with simulation time in Fig. \ref{Fig:05Dt_E}, including the parallel $d_{\parallel}^2 = \langle(x-x_0)^2\rangle$ (panel a) and perpendicular components $d_{\perp}^2 = \langle(y-y_0)^2\rangle + \langle(z-z_0)^2\rangle$ (panel b), in the absence (dashed lines) and presence (solid lines) of radiative losses, based on the power-law injection from Group $\mathcal{A}$. 

It can be seen that the overall trend of the time evolution for $d_{\parallel}^2$ and $d_{\perp}^2$ is the same, due to the same Alfv\'en Mach number $M_{\rm A}=0.51$ and the anisotropic diffusion model (see Eq. (\ref{eq:DMa})). For the GeV (blue lines), TeV (green lines) bands, as well as the wide energy band from $10^1$ to $10^7$ GeV (red lines; which overlapped with the blue lines), the ensemble-averaged squared deviations are proportional to the simulation time ($\propto t$) before $t\sim 10^3$ yr, as expected for diffusion motion of particles (e.g., \citealt{Beresnyak2011}). As seen in this figure, there are a few differences in the time evolution of the deviations between the non-loss and loss cases in the low/wide energy band. This is because of the weak influences of the loss processes in this case (see also details in panels (b) and (d) of Fig. \ref{Fig:01Ur_UE}), and of the energy-dependent parallel diffusion (see Eq. (\ref{eq:Dpar})). Regarding the PeV band, in the case of the non-loss effect, the deviations are proportional to the simulation time ($\propto t$; orange-dashed line) before $t\gtrsim 10^{3}$ yr, which corresponds to the normal diffusion type. While in the case of including radiative cooling, before $t\gtrsim 10^{3}$ yr, the deviations are proportional to $t^{4/5}$ (orange-solid line), corresponding to a sub-diffusion process (\citealt{Lazarian2014, Hu2022}). This implies that the PeV observations originate from the CR electrons suffering sub-diffusion processes.

\begin{figure}[t]
\centering   \includegraphics[width=0.48\textwidth,height=0.22\textheight]{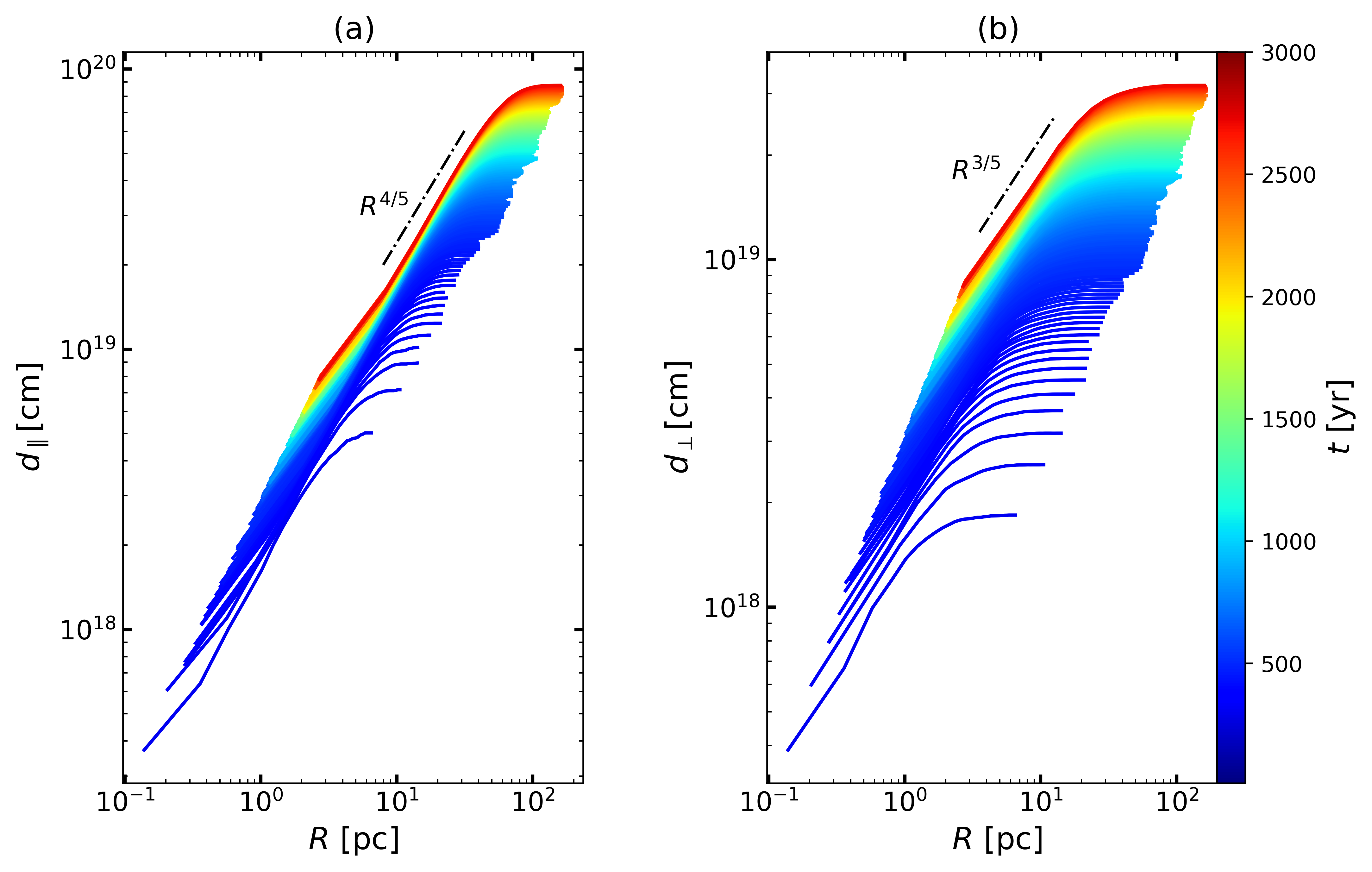}   
\caption{The root mean square of the parallel deviations of particles (panel a) and the perpendicular one (panel b) as a function of the diffusion radius $R$, plotted the non-loss case of $E=$ 1 TeV with $\gamma=5/3$ and $M_{\rm A}=0.51$, where the color bar shows the simulation time in units of yr.}
\label{Fig:06Dr_monoE}%
\end{figure}

By binning particles into different diffusion radii $R = \sqrt{x^2 + y^2 + z^2}$ at each simulation snapshot, we show the time-dependent evolution of the root mean square of the deviations of particles over $R$ in Fig. \ref{Fig:06Dr_monoE} (taking the non-loss case of $E=1$ TeV with $\gamma=5/3$ and $M_{\rm A}=0.51$ as an example), including the parallel (panel a) and perpendicular components (panel b). It can be seen that $d_{\parallel}$ and $d_{\perp}$ present the power-law relationships of $d_{\parallel} \propto R^{4/5}$ (panel a) and $d_{\perp} \propto R^{3/5}$ (panel b), respectively, following a plateau phase at large $R$. At small $R$, the diffusion is slow, while it goes faster with increasing $R$, in agreement with slow diffusion near the source and fast/normal-diffusion diffusion far from the source. For the shallower power law of $d_{\perp} \propto R^{3/5}$, it indicates that the spatial diffusion rate in the perpendicular direction is slower than that in the parallel direction. At the same time, the increment of $d_{\parallel}$ is larger than that of $d_{\perp}$, related to the anisotropic diffusion (i.e., Eq. (\ref{eq:DMa})). Furthermore, with time evolving, the diffusion radius and the spatial extent become larger, for both the parallel and perpendicular diffusion. 

In general, the turbulent cascade happens by interacting with eddies in the direction perpendicular to the local magnetic field, and the energy is transferred from large scales to small scales. It is demonstrated that the level of MHD turbulence tends to become stronger during the cascade (\citealt{Beresnyak2019}). While the eddies are stretched along the parallel direction, showing the $l_{\parallel}\sim l_{\perp}^{2/3}$ anisotropy of turbulence (GS95), where $l_{\parallel}$ and $l_{\perp}$ are parallel and perpendicular scales of the eddy, respectively. Due to the interaction of particles with magnetic turbulence, they will suffer an anisotropic diffusion with the diffusion scales in the parallel direction larger than the perpendicular one.

\subsection{Characterization of the power-law relation} \label{sec:numre.aniso.power-law}

\begin{figure*}[t]
\centering
\includegraphics[width=0.48\textwidth,height=0.24\textheight]{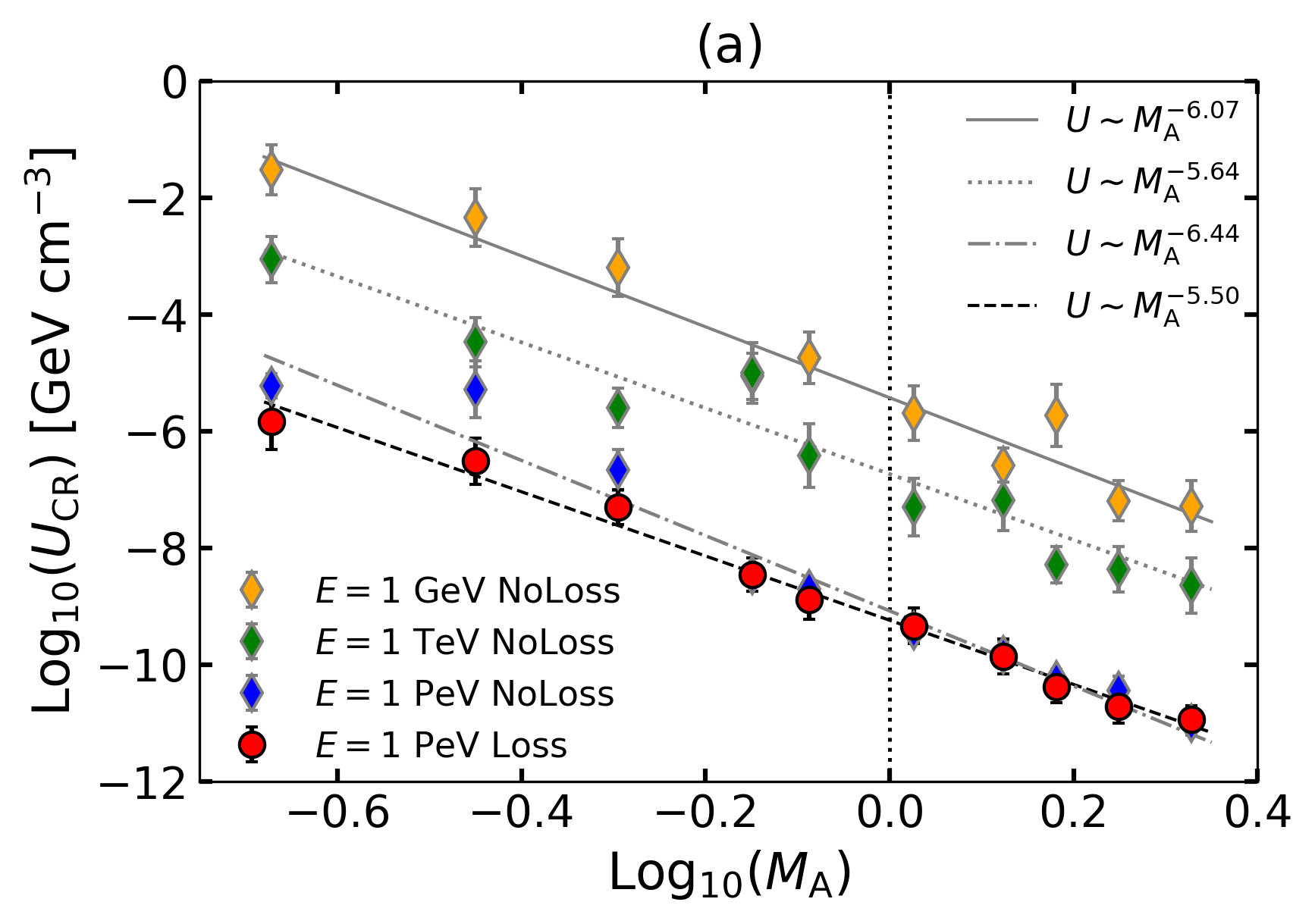}
\includegraphics[width=0.48\textwidth,height=0.24\textheight]{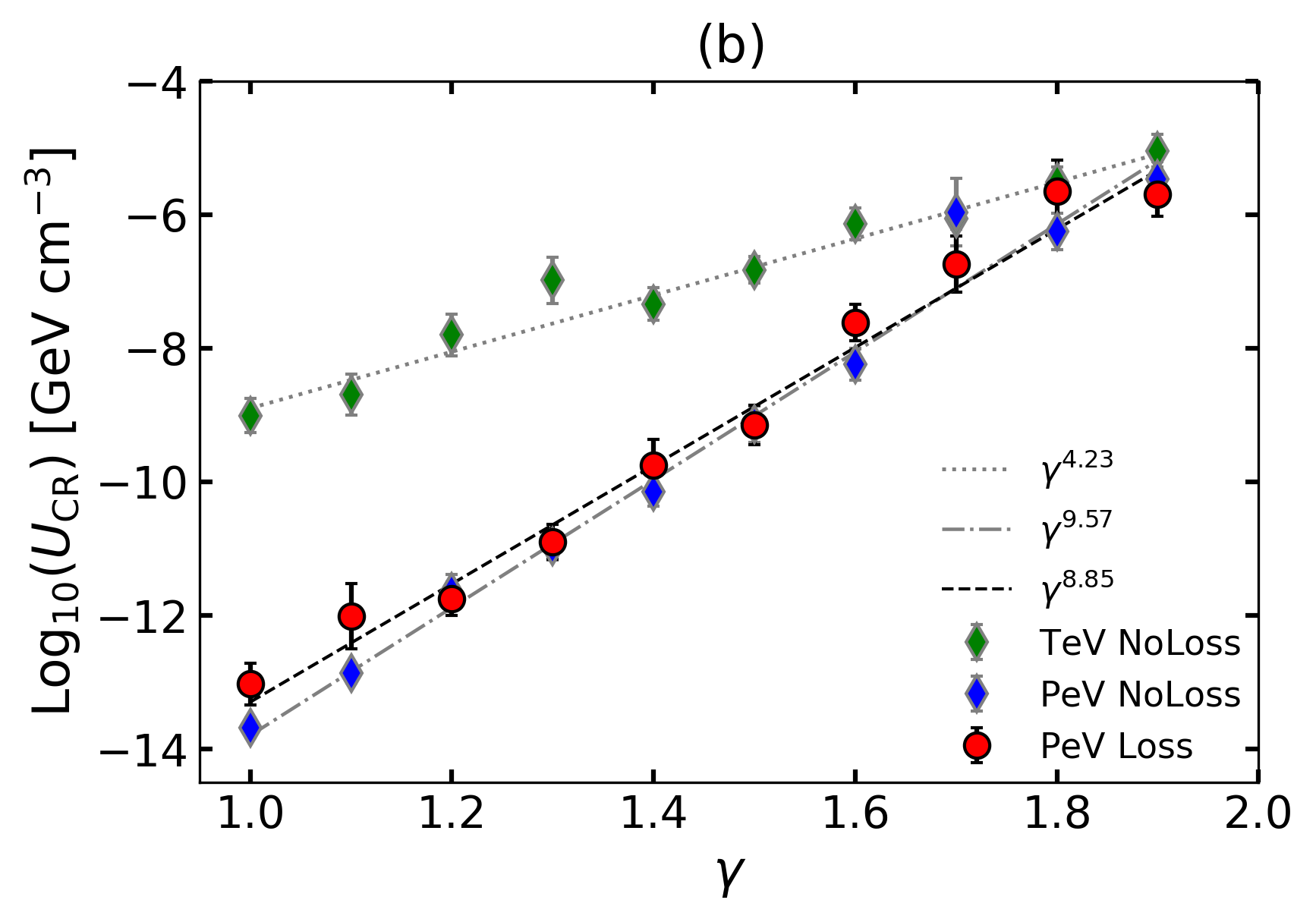}
\includegraphics[width=0.48\textwidth,height=0.24\textheight]{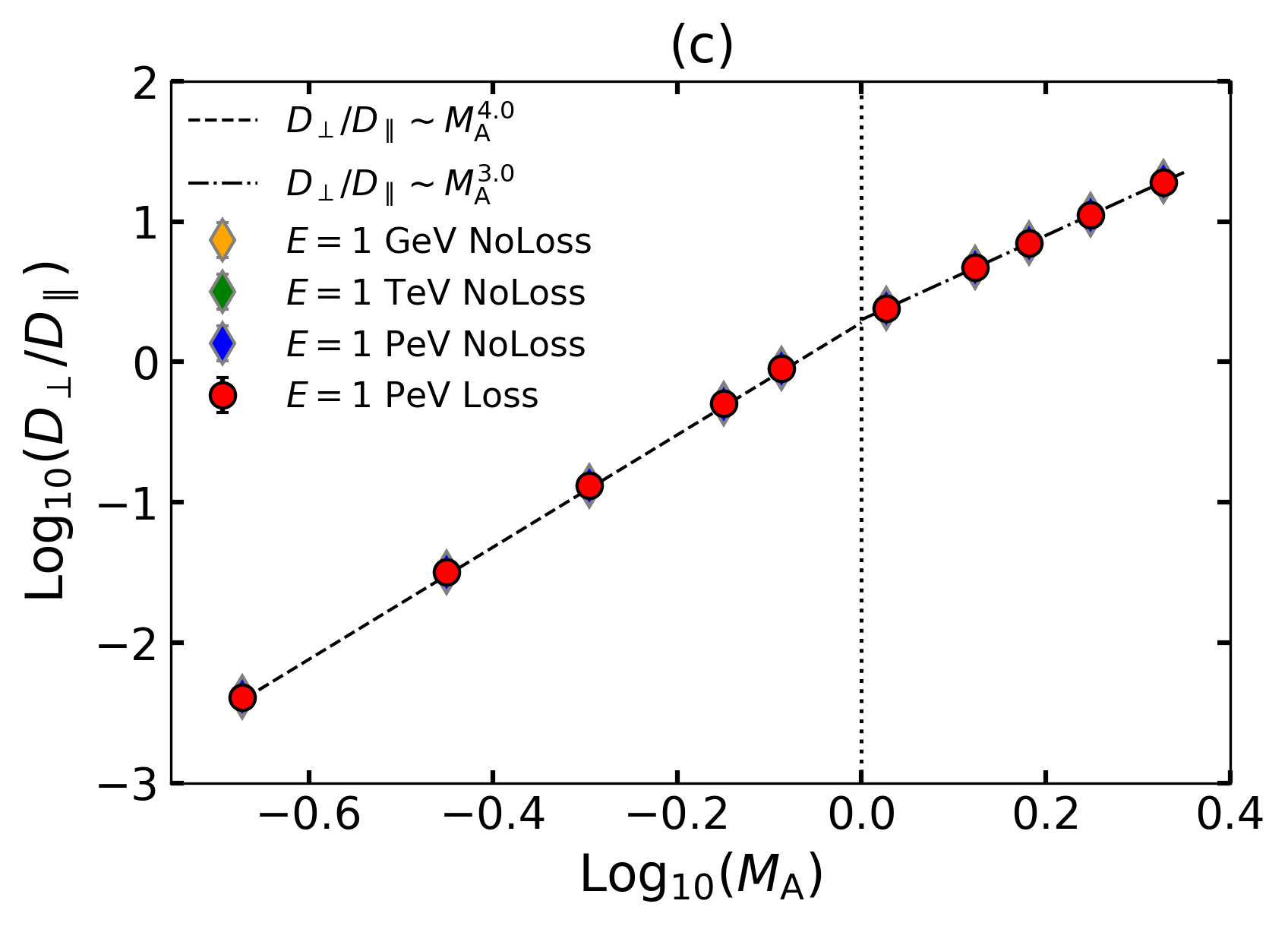}
\includegraphics[width=0.48\textwidth,height=0.24\textheight]{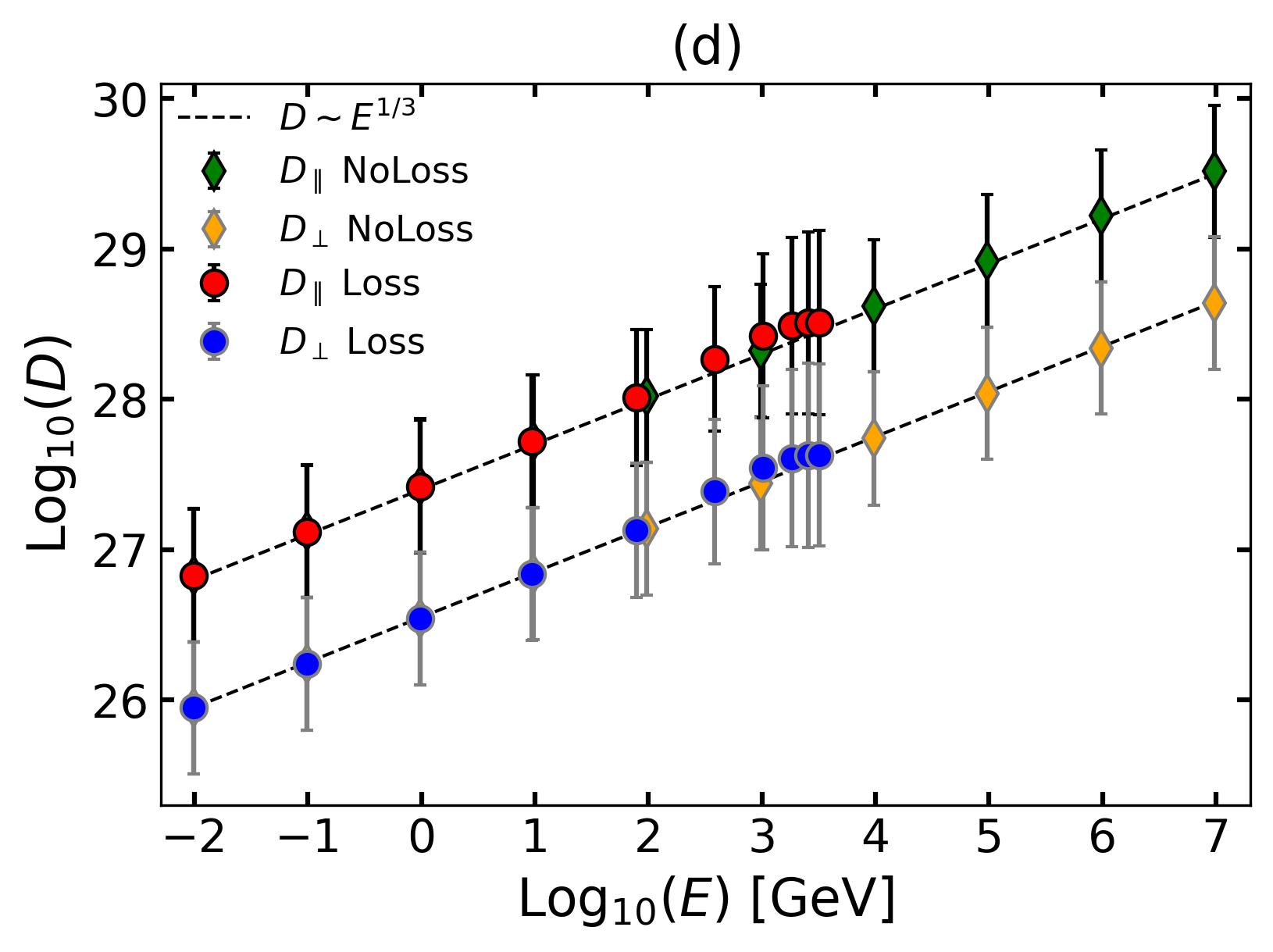}
\caption{Upper row: the averaged energy density $U_{\rm CR}$ as a function of the magnetization parameter $M_{\rm A}$ (panel a) and diffusion index $\gamma$ (panel b). Lower row: the ratio of the perpendicular diffusion coefficient to the parallel one, $D_{\perp}/D_{\parallel}$, vs. $M_{\rm A}$ (panel c); the parallel and perpendicular diffusion coefficients as a function of the averaged energy of particles, considering the cases with (circles) and without (thin-diamonds) radiative losses (panel d). The vertical dotted lines plotted in panels (a) and (c) correspond to $M_{\rm A} = 1$. }
\label{Fig:07UD_MA-Delta}%
\end{figure*}

Given that the relation of $U_{\rm CR}$ vs. $r$ at different magnetization parameters $M_{\rm A}$ presents a continuous change as shown in Fig. \ref{Fig:02Ur_MA-Delta}(a), we speculate that the dependence of $U_{\rm CR}$ on $M_{\rm A}$ may present a correlation. Therefore, we average the value of $U_{\rm CR}$ for each $M_{\rm A}$ to check the possible linear correlation between the averaged $U_{\rm CR}$ and $M_{\rm A}$. The finding of $U_{\rm CR}\propto M_{\rm A}^{-\beta}$, where the index $\beta$ is in the range of [5.50, 6.44], is provided in Fig. \ref{Fig:07UD_MA-Delta}(a) for different mono-energy injection with/without radiative losses. These power-law relationships can be approximated to $U_{\rm CR}\sim M_{\rm A}^{-6.0}$, which is nearly independent of energy and not affected by the radiative losses. The error bars plotted in Fig. \ref{Fig:07UD_MA-Delta} are obtained by the standard deviation. 

Similarly, we study the correlation between the averaged $U_{\rm CR}$ and turbulence index $\gamma$. Adopting the mono-energy injection with/without radiative losses, Fig. \ref{Fig:07UD_MA-Delta}(b) shows the relation of $U_{\rm CR}\propto \gamma^\varepsilon$, where the fitting index $\varepsilon=0$ (GeV; not shown), 4.23 (TeV; without losses), 9.57 (PeV; without losses), and 8.85 (PeV; with losses). On the one hand, the relation between the averaged $U_{\rm CR}$ and $\gamma$ strongly depends on the energy of particles, that is, the higher the energy is, the steeper the fitting index. On the other hand, radiation losses tend to slightly weaken the correlation between the averaged $U_{\rm CR}$ and $\gamma$.

Moreover, we want to explore to what extent the important power-law relations, i.e., Eqs. (\ref{eq:Dpar}) and (\ref{eq:DMa}), associated with MHD turbulence properties, could maintain its own characteristics during the evolution process. We plot the ratio of the perpendicular diffusion coefficient to the parallel one $D_{\perp}/D_{\parallel}$ as a function of the magnetization parameter $M_{\rm A}$ (Fig. \ref{Fig:07UD_MA-Delta}(c)), and the spatial diffusion (including $D_{\parallel}$ and $D_{\perp}$) as the averaged energy $E$ (Fig. \ref{Fig:07UD_MA-Delta}(d)){\footnote{Note that we numerically calculate the parallel and perpendicular coefficient via $D_{\parallel}=d_{\parallel}^2 / 2(t-t_0)$ and $D_{\perp}=d_{\perp}^2 / 2(t-t_0)$, respectively, where $t$ and $t_0$ are the evolution time and initial time, respectively.}}, where the parameter $E$ denotes the energy averaged over both the evolution time and the number of particles in the evolution domain. As seen in Fig. \ref{Fig:07UD_MA-Delta}(c), the ratio $D_{\perp}/D_{\parallel}$ is related to the magnetization parameter $M_{\rm A}$ by the power-law relations of $D_{\perp}/D_{\parallel} \sim M_{\rm A}^{4.0}$ in the sub-Alfv\'enic regime, and $D_{\perp}/D_{\parallel} \sim M_{\rm A}^{3.0}$ in the super-Alfv\'enic one. This numerical finding can well match the initial setting of $D_{\perp,0}/D_{\parallel,0} \sim M_{\rm A}^{\alpha}$, with the index $\alpha=$ 4 and 3. We demonstrate that during the long evolution time, the relation between the ratio $D_{\perp}/D_{\parallel}$ and $M_{\rm A}$ is independent of both radiative losses and kinetic energy of particles, suggesting that this relationship $D_{\perp}/D_{\parallel} \sim M_{\rm A}^{\alpha}$ is an inherent property of particle transport in the magnetic turbulent environment. This implies that the anisotropy properties of MHD turbulence exclusively determine the anisotropic diffusion of CRs.

As for Fig. \ref{Fig:07UD_MA-Delta}(d), we use the mono-energy injection of Group $\mathcal{A}$ with the fixed $\gamma$ = 5/3 and $M_{\rm A}$ = 0.51, considering ten sets of the injection energy of particles range from $10^{-2}$ to $10^{7}$ GeV with an interval of $10$. For the non-loss case, the parallel diffusion coefficient $D_{\parallel}$ (green thin-diamonds) and the perpendicular one $D_{\perp}$ (orange thin-diamonds) both present well a power-law relation of $D_{\perp,\parallel} \sim E^{\delta} = E^{2-\gamma} = E^{1/3}$. While for the case with losses, the energy losses of high energy particles are more serious. We can see from this panel that the average energy distributed in the TeV region after the evolution of particles with energies above TeV. Even considering the radiative losses, $D_{\parallel}$ (red circles) and $D_{\perp}$ (blue circles) still present well with the power-law relation of $D_{\perp,\parallel} \sim E^{1/3}$. It is in good agreement with the expected energy-dependent diffusion process. Consequently, the robust relations of $D_{\perp}/D_{\parallel} \sim M_{\rm A}^{\alpha}$ (see Eq. (\ref{eq:DMa})), and $D_\parallel \sim E^{\delta}$ (see Eq. (\ref{eq:Dpar})) can be applied to understand the TeV and PeV gamma-ray emissions observed and predict the potential observational phenomenon.

\section{Discussion}\label{sec:discu}
Through numerically solving the CR transport equation in a Crab-like nebula environment, we explored the interaction of high-energy CR electrons with magnetic turbulence in the framework of an anisotropic diffusion model. Specifically, regarding the magnetization parameter and turbulence spectral index as the adjustable parameters, we explore their influence on CR spectral, spatial diffusion, and anisotropy morphology.

The propagation process of CRs can be described with the time-dependent Fokker-Planck equation (\citealt{Gaisser1990, Berezinskii1990, Schlickeiser2002}). Provided that one considers different simplifications, the transport equation can be solved analytically (\citealt{Webber1992, Maurin2002, Shibata2004}). However, this simplification process may lead to the loss of CR propagation information. Therefore, the numerical/simulation approach is the main channel for revealing the CR propagation. At present, there are a number of numerical codes developed, e.g., GALPROP (\citealt{Strong1998, Moskalenko1998}), PICARD (\citealt{Kissmann2014}), DRAGON (\citealt{Evoli2008}), CRPROPA (\citealt{Merten2017}), and CRIPTIC (\citealt{Krumholz2022}), to understand CR observations. In the current work, we explored the diffusion properties of CRs using the CRIPTIC code, which is a new code for calculating both CR transport and observable emission. 

In the framework of anisotropic diffusion, we investigated the influence of the magnetization parameter $M_{\rm A}$ and turbulence index $\gamma$ on the CR spatial distributions. In the sub-Alfv\'enic regime, we found that the resulting morphology and size of CR spatial distribution are consistent with that of emission images (\citealt{Liu2019, De2022}) within small viewing angles, while in the case of super-Alfv\'enic turbulence, similar spatial distributions are also formed in the direction perpendicular to the mean magnetic field. Note that the stronger the anisotropy, the larger $M_{\rm A}$ is, and the smaller the spatial extent, the larger $\gamma$. Therefore, when changing the viewing angle with both the large $M_{\rm A}$ and $\gamma$, the expected spatial structures can be reproduced. For instance, with $D_{\perp}/D_{\parallel} \sim M_{\rm A}^{3}$ in the super-Alfv\'enic turbulence\footnote{This work mainly focuses on the strong turbulence region, that is, the spatial scale is smaller than the transition scale $l_{\rm A}$. For instance, considering the magnetic field strength of 50 $\mu$G (corresponding to the super-Alfv\'enic regime) and the maximal energy of CRs of 10 PeV, we can simply estimate the maximal gyroradius of CRs with a velocity of $v_{\rm e}$ as (\citealt{Barreto-Mota2024}) $R_{\rm g, max} \approx \frac{(E/1\ {\rm PeV})}{(B/1\ \mu{G})} \frac{v_{\rm e}}{c} \approx \frac{(10\ {\rm PeV}/1\ {\rm PeV})}{(50\ \mu{\rm G}/1\ \mu{\rm G})} \approx 0.2\ {\rm pc}$, which is much smaller than the parallel diffusion scales $l_{\perp} \approx$ hundreds of pc, as seen in Fig. \ref{Fig:04Anisotropy_MA-Delta}(c) for one super-Alfv\'enic case. Given such a small gyroradius, CR diffusion is strongly subjected to the anisotropy of magnetic fields.}, when $M_{\rm A} \gtrsim 4.7$, the parallel diffusion coefficient is about hundreds of times lower than the perpendicular one. This provides an alternative way for explaining the slow diffusion and morphology of observed TeV halos, with the LOS approximately aligned with the mean magnetic field. 

For the other plausible explanation of the slow diffusion, i.e., the mirror diffusion, \cite{Lazarian2021} analytically predicted the mirror diffusion is generally slower than the diffusion of non-bouncing CRs with small pitch angles between the local mean magnetic field and velocity of particles that undergo gyroresonant scattering. Subsequently, many numerical studies focused on exploring the diffusion and acceleration properties of CRs in the frame of mirror diffusion (e.g., \citealt{Xu2021, Xu2023, Zhangchao2023, Zhangchao2024, Barreto-Mota2024}). They found that the mirror diffusion of CRs with sufficiently large pitch angles can explain the slow diffusion around the CR source. CRs may stochastically undergo slow mirror and fast scattering diffusion in the Galactic diffuse medium away from CR sources. We also expect that the mirror diffusion may be a plausible explanation for the slow diffusion near the source, which needs further research. 

In the current work, in addition to the continuous injection used throughout this paper, we also tested the case of instantaneous, finding that CR's overall evolution is similar to the continuous one. The only difference is the lack of the power-law relationship between the energy density and the effective radius. We also tested the streaming interaction for a wide distribution of CR energy (up to 10 PeV) and found that the streaming interaction almost does not change our results. We note that the streaming affects only the diffusion of the GeV CRs (see also the review \cite{Cesarsky1980} and \cite{Farmer2004} for the theoretical expectation). Therefore, the results we presented in the current work do not consider the streaming interaction. Since our study focuses on the CR electrons, it is enough to take the cooling effects of synchrotron radiation and IC scattering into account. When studying diffusion processes of the CR protons, one has to consider more cooling effects such as the coulomb, ionization, bremsstrahlung, and positron annihilation. Considering the PWN environment as a general research object, this work mainly focused on understanding how CR electrons interact with turbulent magnetic fields. Studies explaining real observations will be discussed elsewhere.

\section{Summary}\label{sec:summary}
Considering PWN as a general research case in this paper, we numerically study the interactions of CRs with the ambient turbulent magnetic field. Focusing on the CR spectral distribution, anisotropy morphology, and spatial diffusion coefficients, our work is devoted to understanding CR propagation processes in the presence of magnetic turbulence. The main findings are briefly summarized as follows: 
   
\begin{enumerate} 
   \item We find that the energy density $U_{\rm CR}$ without radiative losses displays the power-law relationships of $U_{\rm CR} \propto r^{-6/5}$ and $U_{\rm CR} \propto 1/E$, which are modified as $U_{\rm CR} \propto r^{-11/5}$ and $U_{\rm CR} \propto E^{-8/5}$ in the presence of radiative losses, respectively.

   \item The morphology of the CR spatial distribution strongly depends on the properties of turbulence. Namely, the magnetization parameter $M_{\rm A}$ affects the spatial anisotropic distribution, while the turbulence spectral index $\gamma$
   its spatial extent. Note that the viewing angle also affects the spatial anisotropic distribution, consistent with the previous studies (e.g., \citealt{Liu2019,De2022}).
   
   \item The CR electrons suffer from a slow diffusion near the source and a fast/normal diffusion away from the source. The effect of radiative losses, which happen far from the source, can suppress CR diffusion processes, resulting in a sub-diffusion behavior. 
   
   \item The averaged energy density $U_{\rm CR}$ distributes a power-law relationship of $U_{\rm CR} \propto M_{\rm A}^{-6.0 \pm 0.5}$, independent of both CR kinetic energy and radiative losses.

   \item Our simulation demonstrates that the relations of $D_{\parallel} \sim E^{2-\gamma}$ and $D_{\perp}/D_{\parallel} \sim M_{\rm A}^{\alpha}$ maintain their features throughout a long evolution period. Since these two relations establish a robust connection between turbulent magnetic fields and CR diffusion coefficients, one can apply them to understand complex astrophysical processes related to turbulence cascades.   
   
\end{enumerate}

\begin{acknowledgements}
We appreciate Mark R. Krumholz's helpful discussions. The authors thank the support from the National Natural Science Foundation of China (grant No. 12473046). J.F.Z. also thanks the Hunan Natural Science Foundation for Distinguished Young Scholars (No. 2023JJ10039) and the China Scholarship Council for overseas research funds.

\end{acknowledgements}

\bibliographystyle{aa}
\bibliography{aa}

\begin{appendix}
\section{The main procedures of our simulation experiments}\label{sec:appendixA}
\cite{Lazarian2014} analytically predicted the potential relation between the ratio of the perpendicular diffusion coefficient to the parallel component and the magnetization parameter of
\begin{equation}
  D_{\perp}/D_{\parallel} \approx M_{\rm A}^{\alpha} , \label{eq:A1DMa}
\end{equation}
in the strong turbulence range, where $\alpha=$ 4.0 for the sub-Alfv\'enic turbulence and $\alpha=$ 3.0 for the super-Alfv\'enic case. This relation has been verified by the test-particle simulations (\citealt{Xu2013, Maiti2022}). However, analytical and numerical studies did not consider the possible effect of both the change of CR energies and their radiative losses on the power-law relation. We note that some works, such as \cite{Liu2019} and \cite{De2022}, directly used the relation of $D_{\perp}=D_{\parallel}M_{\rm A}^4$ to explain the slow diffusion phenomena. One of the motivations of the current work is to explore whether the relationship between the ratio of the spatial diffusion coefficient and the magnetization parameter can maintain its properties during the propagation process of CRs. 

To conduct our simulations using the CRIPPTIC code, we need to obtain the input parameters that the CRIPPTIC code can invoke, by setting some physical parameters related to MHD turbulence dynamics and CRs. For an example of our simulations, which considers the case of $E\in 10^{[1,\ 7]}$ GeV with radiative loss processes (including synchrotron radiative losses due to the presence of the ISM magnetic field and IC scattering losses due to both interstellar radiation field with a temperature of 20 K and cosmic microwave one in the Klein-Nishina regime), we describe the main procedures as follows.

$\bm{{\rm Step\ 1}}$: Since CRIPTIC takes the sub-grid values for diffusion coefficients as input parameters, $D_{\parallel}$ needs to be taken as the baseline value. With the energy-dependent parallel diffusion coefficient (\citealt{Seo1994, Trotta2011, Dempers2020})
\begin{equation}
  D_{\parallel} = D_0(E/m_{\rm e}c^2)^{\delta} \,, \label{eq:A2Dpar}
\end{equation}
where $m_{\rm e}$ and $c$ are the mass of the electron and light speed, respectively, we first obtain the initial parallel diffusion coefficient $D_{\parallel,0}$. Note that the CRIPTIC code considers the normalized spatial diffusion coefficient $D_0$, diffusion index $\delta$, and the kinetic energy of electrons $E$ as input parameters. In this case, we fixed $D_0 = 1.0\times 10^{28}\ {\rm cm^{2}\ s^{-1}}$ for ISM diffusion (\citealt{Heesen2019}) and considered a power-law energy distribution of $d\dot{n}/dE \propto E^{-q}$ with $q=2.2$. By setting the parameter $\delta$, we can obtain a connection between the diffusion index $\delta$ and magnetic turbulence index $\gamma$ (for instance, setting $\delta=1/3$, we have $\gamma=5/3$ corresponding to the Kolmogorov-type turbulence; see more details in Sect. \ref{sec:theor.diffturbulence}).

$\bm{{\rm Step\ 2}}$: On the basis of the established $D_{\parallel,0}$ in Step 1, we can constrain the initial perpendicular diffusion coefficient $D_{\perp, 0}$ via \ref{eq:A1DMa}. We note that the CRIPTIC code parameterizes the ratio of $D_{\perp}/D_{\parallel}$ as $\chi$. In this work, we define $\chi = M_{\rm A}^{\alpha}$ to establish the link between the ratio of $D_{\perp}/D_{\parallel}$ and the magnetization parameter $M_{\rm A}$. Given theoretical expectation and numerical confirmation of the index $\alpha$ value mentioned above, we fixed $\alpha=$ 4.0 for the sub-Alfv\'enic turbulence and $\alpha=$ 3.0 for the super-Alfv\'enic one. For the magnetization parameter, i.e., the Alfvén Mach number $M_{\rm A} = v/v_{\rm A} = v/(B_0/\sqrt{4\pi \rho})$, we controlled its value by regulating the mean magnetic field strength $B_0$ along with the $x$-axis direction, considering a uniform region containing gas with a density of $\rho \simeq 10^{-24}\ {\rm g\ cm^{-3}}$ and velocity of $v \simeq 300\ {\rm km\ s^{-1}}$. 

$\bm{{\rm Step\ 3}}$: Furthermore, we use the established parallel diffusion coefficient $D_{\parallel,0}$ in Step 1 to obtain the initial energy diffusion coefficient $D_{\rm EE,0}$ by $D_{\rm EE} = E^{2}v_{\rm A}^{2}/9D_{\parallel}$ (\citealt{Michalek1996}).

Based on this framework, we explore the interaction of high-energy CR electrons with magnetic turbulence, regarding the magnetization parameter $M_{\rm A}$ (in the range of [0, 3]), turbulence spectral index $\gamma$ ($\gamma=2-\delta$ in the range of [1.0, 1.9], i.e., $\delta \in$ [0.1, 1.0]), and particle’s energy $E$ (in the range of [$10^{-2}$, $10^7$] GeV) as free parameters, together with the radiative processes, and divide our simulations into three groups. The detailed parameters of each group are provided in Sect. \ref{sec:numre.Model}.

\end{appendix}

\end{document}